\documentclass[prd,twocolumn,showpacs,showkeys,preprintnumbers,nofootinbib,amsmath,eqsecnum,superscriptaddress]{revtex4}

\usepackage[dvips,final]{graphicx}
 \usepackage{amssymb}
  \usepackage{amsmath}
   \usepackage{amsfonts}
    \usepackage{epsfig}
     \usepackage{bm}
      \usepackage{multirow}
       \usepackage{tabularx}

\newcommand{\A}{{\scriptscriptstyle A}}

\newcommand{\sL}{{\scriptscriptstyle L}}
\newcommand{\R}{{\scriptscriptstyle R}}

\newcommand{\W}{{\scriptscriptstyle W}}

\newcommand{\sTC}{{\scriptscriptstyle TC}}

\begin{document}

\preprint{LU-TP 14-21}

\title{Composite scalar Dark Matter from vector-like $SU(2)$ confinement}

\author{Roman Pasechnik}
\affiliation{Department of Astronomy and Theoretical Physics, Lund
University, SE-223 62 Lund, Sweden}

\author{Vitaly Beylin}
\affiliation{Research Institute of Physics, Southern Federal
University, 344090 Rostov-on-Don, Russia}

\author{Vladimir Kuksa}
\affiliation{Research Institute of Physics, Southern Federal
University, 344090 Rostov-on-Don, Russia}

\author{Grigory Vereshkov}
\affiliation{Research Institute of Physics, Southern Federal
University, 344090 Rostov-on-Don, Russia} \affiliation{Institute for
Nuclear Research of Russian Academy of Sciences, 117312 Moscow,
Russia}

\begin{abstract}
A toy-model with $SU(2)_{\rm TC}$ dynamics confined at high scales 
$\Lambda_{\rm TC}\gg 100$ GeV enables to construct Dirac UV completion 
from the original chiral multiplets predicting a vector-like nature of 
their weak interactions consistent with electroweak precision tests. 
In this work, we investigate a potential of the lightest scalar baryon-like (T-baryon) state 
$B^0=UD$ with mass $m_B\gtrsim 1$ TeV predicted by the simplest two-flavor 
vector-like confinement model as a Dark Matter (DM) candidate. We show that two 
different scenarios with the T-baryon relic abundance formation before and 
after the electroweak (EW) phase transition epoch lead to symmetric (or mixed) 
and asymmetric DM, respectively. Such a DM candidate evades existing direct 
DM detection constraints since its vector coupling to $Z$ boson absents at tree level, 
while one-loop gauge boson mediated contribution is shown to be vanishingly small close 
to the threshold. The dominating spin-independent (SI) T-baryon--nucleon 
scattering goes via tree-level Higgs boson exchange in the $t$-channel. 
The corresponding bound on the effective T-baryon--Higgs coupling has 
been extracted from the recent LUX data and turns out to be consistent with naive 
expectations from the light technipion case $m_{\tilde \pi}\ll \Lambda_{\rm TC}$.
The latter provides the most stringent phenomenological constraint on 
strongly-coupled $SU(2)_{\rm TC}$ dynamics so far. Future prospects 
for direct and indirect scalar T-baryon DM searches in astrophysics 
as well as in collider measurements have been discussed.
\end{abstract}



\maketitle

\section{Introduction}

After recent discovery of the Higgs boson \cite{Aad:2012tfa,Chatrchyan:2012ufa} 
at the LHC and follow-up precision studies of its interactions with known matter
\cite{SM-tests}, a rough picture of consistency with the Standard Model
(SM) has began to emerge. The SM is being widely perceived as an effective 
field theory providing quite accurate results and matches the physical reality 
at low energies to a rather high precision. So the observed consistency
does not mean yet that the nature of the Higgs boson and electroweak
symmetry breaking (EWSB) is completely understood at the fundamental level. 
One of the immediate questions that challenge our current understanding of
symmetries in Nature is what initiates the Higgs mechanism of 
electroweak symmetry breaking (EWSB) in the SM. Namely, whether it
is actually SM-like or in fact an effective description appearing as
a consequence of a new fundamental symmetry. The experimental
precision in the Higgs SM sector should be noticeably increased to the level of precision
tests achieved in other SM sectors (i.e. flavor and electroweak (EW) gauge boson sectors) 
in order to draw final conclusions. In particular, in the EW sector of the SM the
agreement of theoretical predictions with precision measurements 
takes place at the level of $\sim 10^{-3}$ in relative error which 
remains a rather big challenge for ongoing experimental investigations 
of the SM Higgs sector at the LHC.

A new strongly-coupled dynamics at a TeV energy scale is often
considered to be responsible for EWSB in the SM \cite{TC-1,TC-2,Extended-TC-1,Extended-TC-2}. 
Namely, it initiates the EWSB dynamically by means of strongly-interacting
techniquark (or T-quark) condensation at low energy scales $\sim 100$ GeV. A straightforward
analogy to this effect is the spontaneous chiral symmetry breaking
in non-perturbative chiral QCD with only difference that the
effective Higgs mechanism is initiated by a non-diagonal T-quark
condensate. Such a new dynamics unavoidably predicts a plenty of new
states, most importantly, composite Higgs-like particles 
\cite{Vecchi:2013bja,Barducci:2013wjc} and partners of 
SM fermions \cite{DeSimone:2012fs}, whose properties depend on
the group-theoretical structure of underlined theory and its
ultraviolet (UV) completion. Besides the composite Higgs-like state,
the latter predict a plenty of new relatively light pseudo-Goldstone
composite states, like technipions (or T-pions below), techni-$\eta$, techni-$K$, etc,
whose search in low invariant mass regions is strongly
limited by large SM backgrounds and weak couplings to SM particles. 
The search for a family of new light (pseudo)scalar states with 
invariant masses below $200$ GeV is one of the priority topics
for new strongly-coupled physics searches at the LHC Run II.

A large number of various realisations of such a new dynamics at a TeV
scale, commonly dubbed as ``Technicolor'' (TC) or ``compositeness''
scenarios, have been proposed in the literature so far 
(for a review, see e.g. Refs.~\cite{Hill:2002ap,Sannino}).
Such a big variety, however, has got reduced by severe
electroweak (EW) precision tests \cite{Peskin-1,Peskin-2} and recent SM Higgs-like
particle observations. At present, among the most appealing scenarios 
of dynamical EWSB consistent with current constraints is a class 
of models with vector-like (Dirac) UV completion -- 
the vector-like confinement (VLC) scenario. The phenomenological importance
of the vector-like confinement has been broadly discussed in e.g. Ref.~\cite{Kilic:2009mi}
without referring to its implication to the dynamical EWSB.
The simplest realisation of the VLC scenario of the EWSB with two vector-like 
or Dirac techniflavors and a SM-like Higgs boson has been studied in 
Refs.~\cite{Pasechnik:2013bxa,Pasechnik:2013kya,Lebiedowicz:2013fta} and 
very recently has emerged in composite Higgs scenarios with 
confined $SU(2)_{\rm TC}$ \cite{Cacciapaglia:2014uja,Hietanen:2014xca}. 
In this paper, we discuss implications of the vector-like confinement
to Dark Matter (DM) astrophysics.

Besides the dynamical nature of EWSB in the SM and a possible
compositeness of the Higgs boson, another very important prediction
of QCD-like TC scenarios based upon $SU(N_{\rm TC})_{\rm TC}$ 
confined symmetry is the existence of heavy composite
baryon-like states possessing an additional conserved quantum number.
The lightest neutral technibaryon (or T-baryon) state
thus appears to be stable and weakly interacting with ordinary
matter. If a new strong dynamics exists in Nature just above the EW
scale $M_{\rm EW}\simeq 200$ GeV and if there is a mechanism for
T-baryon asymmetry generation analogical to that of baryon
asymmetry, such new particles could be abundantly produced in early
Universe and survived until today in the form of DM
\cite{Nussinov}. These ideas is widely discussed in the literature 
during past two decades. 

So far, a number of different models of composite DM candidates and hypotheses 
about their origin and interactions has been proposed. Generic DM signatures from
TC-based models with stable T-baryons were discussed e.g.
in Refs.~\cite{Gudnason-1,Gudnason-2,Khlopov-1,Khlopov-2,DelNobile:2011uf} (for a review see
also Ref.~\cite{Sannino:2009za} and references therein). In particular, 
well-known minimal dynamical EWSB mechanisms predict relatively light 
T-baryon states as pseudo Nambu-Goldstone bosons of
the underlying gauge theory \cite{Belyaev:2010kp,Ryttov,Lewis:2011zb}. The latter
can naturally provide partially-asymmetric or asymmetric DM (ADM) 
candidates if one assumes the existence of a T-baryon asymmetry in Nature similarly to ordinary
baryon asymmetry \cite{Belyaev:2010kp,Lewis:2011zb} (for a review on ADM models, 
see e.g. Ref.~\cite{Petraki:2013wwa} and references therein). Having similar mechanisms
for ordinary matter and DM formation in early Universe one would
expect the DM density to be of the same order of magnitude as that
of baryons. Depending on a particular realization of dynamical EWSB
mechanism such composite DM candidates may be self-interacting which
helps in avoiding problematic cusp-like DM halo profiles
\cite{Kouvaris:2013gya}. The ongoing search for the DM 
in both direct and indirect measurements can thus
provide further tight constraints on possible TC scenarios
additional to those coming from the LHC.

To this end, in Ref.~\cite{Pasechnik:2013kya} it has been demonstrated explicitly that 
the TC scenarios with an odd confined $SU(2n+1)_{\rm TC},\,n=1,2,\dots$
symmetry are most likely ruled out by recent constraints on the 
spin-independent DM-nucleon scattering cross section \cite{Aprile:2012nq,Akerib:2013tjd}. 
In particular, stable Dirac T-neutron DM predicted by the confined 
QCD-like $SU(3)_{\rm TC}$ symmetry is excluded due to its large 
tree-level vector gauge coupling to the $Z$ boson unless it is not directly 
coupled to weak isospin $SU(2)_{\rm W}$ sector, only via a small mixing. 

However, confined even $SU(2n)_{\rm TC},\,n=1,2,\dots$ symmetries 
giving rise to scalar T-baryon $B=QQ$ (diquark-like) states instead are 
void of this problem. Indeed, the elastic scattering of scalar T-baryons off
nucleons occurs mainly via the Higgs boson exchange at tree level
and is strongly suppressed compared to stable Dirac composites. 
As was advocated recently in Refs.~\cite{Lewis:2011zb,Appelquist:2014dja} 
the {\it light} scalar T-baryons (or T-diquarks) can play a role of pseudo-Goldstone 
bosons under global $SU(4)$ symmetry such that the lightest neutral $UD$ T-baryon 
state could become a new appealing composite asymmetric or mixed DM candidate.
In this Letter, we are focused primarily on a simpler {\it global chiral} 
$SU(2)_L\otimes SU(2)_R$ symmetry acting on complex vector-like (Dirac) UV completion 
in {\it local gauge} $U(1)_{\rm Y}\otimes SU(2)_{\rm W}\otimes SU(2)_{\rm TC}$ 
symmetry which possesses an additional conserved T-baryon charge 
\cite{Pasechnik:2013bxa,Pasechnik:2013kya,Lebiedowicz:2013fta}.
Continuing earlier line of studies, now we discuss important phenomenological implications
of {\it heavy} scalar T-baryons $m_B\gtrsim 1$ TeV for direct DM searches 
in astrophysics and collider measurements. Specifically, we demonstrate that 
the heavy scalar T-baryon is a good candidate for self-interacting {\it symmetric} 
DM which is within a projected few-year reach at direct detection experiments.

\section{Scalar T-baryon interactions}

\subsection{Vector-like confinement and Dirac T-quarks}
\label{sec:VLC}

In the considering version of the model, the large scalar T-baryon mass terms explicitly break global chiral $SU(4)$,
so in this case it suffices to work within the global chiral $SU(2)_{\rm R}\otimes SU(2)_{\rm L}$ symmetry 
which classifies the lightest TC states only, similarly to that in hadron physics. This in variance
with models of Refs.~\cite{Lewis:2011zb,Appelquist:2014dja} where the pseudo-Goldstone T-baryon states
can be arbitrarily light and are typically considered to be around the electroweak scale.

Consider the simplest vector-like TC model with single $SU(2)_{\rm W}$ doublet of
Dirac T-quarks confined under a new strongly-coupled gauge symmetry 
$SU(N_{\rm TC})_{\rm TC}$ at the T-confinement scale 
$\Lambda_{\rm TC}\gtrsim 1$ TeV
 \begin{eqnarray} \label{Tdoub}
 {\tilde Q} = \left(
      \begin{array}{c}
         U \\
         D
      \end{array}
             \right)\,, \qquad 
Y_{\tilde Q}=\begin{cases}
0,                 & \text{if } N_{\rm TC}=2 \,, \\
1/3,             & \text{if } N_{\rm TC}=3 \,.
\end{cases}
 \end{eqnarray}
where the T-quark doublet hypercharges are chosen to provide
integer-valued electric charges of corresponding bounds states. 
The case of $N_{\rm TC}=3$ has been studied in Refs.~\cite{Pasechnik:2013bxa,Pasechnik:2013kya},
and here we are focused primarily on $N_{\rm TC}=2$ theory where phenomenologically consistent 
vector-like weak interactions of an underlined UV completion (i.e. Dirac T-quarks) can be naturally
obtained from a conventional chiral one (for $N_{\rm TC}=3$ this is not the case).

In order to demonstrate this fact explicitly, let us start with two generations ($A=1,2$) of left-handed 
T-quarks $Q_{\sL(\A)}^{a\alpha}$ transformed under gauge $SU(2)_{\rm W}\otimes SU(2)_{\rm TC}$ as
\begin{equation}
\begin{array}{c}
\displaystyle
 {\tilde Q}^{a\alpha'}_{\sL(\A)}={\tilde Q}^{a\alpha}_{\sL(\A)}+
 \frac{i}{2}g_\W\theta_k\tau^{ab}_k{\tilde Q}^{b\alpha}_{\sL(\A)} \\ 
 \displaystyle +
 \frac{i}{2}g_{\sTC}\varphi_k\tau^{\alpha\beta}_k{\tilde Q}^{a\beta}_{\sL(\A)}\ ,
\end{array}
 \label{3}
 \end{equation}
where $a=1,2$ is the index of fundamental representation of weak 
isospin $SU(2)_{\rm W}$ group, $\alpha=1,2$ in the index
of fundamental representation of T-strong $SU(2)_{\rm TC}$, and $Y_{\tilde Q}=0$.
Now, let us keep the first generation of T-quarks unchanged and 
apply the charge conjugation to the second generation such that
\begin{equation}
\begin{array}{c}
\displaystyle \hat {\bf C}Q^{a\alpha}_{\sL(2)}=Q^{Ca\alpha}_{\sL(2)}\,, \\
\displaystyle
 Q^{Ca\alpha'}_{\sL(2)}=Q^{Ca\alpha}_{\sL(2)}-\frac{i}{2}
 g_\W\theta_k(\tau^{ab}_k)^*Q^{Cb\alpha}_{\sL(2)} \\
\displaystyle
 - \frac{i}{2}g_{\sTC}\varphi_k(\tau^{\alpha\beta}_k)^*Q^{Ca\beta}_{\sL(2)}\ .
\end{array}
 \label{5}
 \end{equation}
The charge conjugation of a chiral fermion changes its chirality. 
This fact enables us to define the corresponding right-handed field as
\begin{equation}
\begin{array}{c}
\displaystyle
 Q^{a\alpha}_{\R(2)}\equiv \varepsilon^{ab}\varepsilon^{\alpha\beta}Q^{Cb\beta}_{\sL(2)}\ , \quad
 \varepsilon^{ab}=\varepsilon^{\alpha\beta}=\left(\begin{matrix}0&1\\-1&0\end{matrix}\right)\ .
\end{array}
 \label{7}
 \end{equation}
Starting from the gauge group transformation property (\ref{5}) and applying $SU(2)$ 
defining relations like $\delta^{ab}=\varepsilon^{ac}\varepsilon^{bc}$ and
\[
\varepsilon^{ab}(\tau^{bc}_k)^*\varepsilon^{cf}=\tau^{af}_k,
\qquad  \varepsilon^{\alpha\beta}(\tau^{\beta\gamma}_k)^*
\varepsilon^{\gamma\mu}=\tau^{\alpha\mu}_k\ ,
\]
it is rather straightforward to show that
\begin{equation}
\begin{array}{c}
\displaystyle
 Q^{a\alpha'}_{\R(2)}=Q^{a\alpha}_{\R(2)}+\frac{i}{2}
 g_\W\theta_k\tau^{ab}_kQ^{b\alpha}_{\R(2)} \\ 
 + \displaystyle
 \frac{i}{2}g_{\sTC}\varphi_k\tau^{\alpha\beta}_kQ^{a\beta}_{\R(2)}\ .
\end{array}
 \label{10}
 \end{equation}

By a comparison of Eq.~(\ref{10}) with Eq.~(\ref{3}) one notices
that the transformation properties of the {\it right-handed} T-quark field
obtained by charge conjugation and transposition of the left-handed
field of the {\it second} generation coincide with the
transformation properties of the {\it left-handed} field of the {\it
first} generation. Therefore, starting initially with two chiral 
(left-handed) T-quark generations we arrive at {\it one vector-like 
generation of (Dirac) T-quarks}, namely
\begin{equation}
\begin{array}{c}
\displaystyle
Q^{a\alpha}=Q^{a\alpha}_{\sL(1)}+Q^{a\alpha}_{\R(2)}=Q^{a\alpha}_{\sL(1)}+
\varepsilon^{ab}\varepsilon^{\alpha\beta}Q^{Cb\beta}_{\sL(2)}\ .
\end{array}
 \label{11}
 \end{equation} 
As was argued for the first time in Ref.~\cite{Pasechnik:2013bxa}, practically 
any simple Dirac UV completion with chirally-symmetric weak interactions easily 
evade the most stringent electroweak constraints which is the basic motivation 
for the VLC scenario.

The phenomenological interactions of the constituent Dirac T-quarks and the 
lightest T-hadrons, namely, the scalar SM-singlet T-sigma $S$ field, and the 
$SU(2)_{\rm W}$-adjoint triplet of T-pion fields $P_a,\,a=1,2,3$, are described 
by the (global) chiral $SU(2)_{\rm R}\otimes SU(2)_{\rm L}$ invariant low-energy 
effective Lagrangian in the linear $\sigma$-model (L$\sigma$M)
 \begin{eqnarray} \nonumber
  && {\cal L}_{\rm L\sigma M} = \frac12\, \partial_{\mu} S\,
  \partial^{\mu} S +
  \frac12 D_{\mu} P_a\, D^{\mu} P_a + i \bar{{\tilde Q}}\hat{D}{\tilde Q} \\
  && -\, g_{\rm TC} \bar{{\tilde Q}}(S+i\gamma_5\tau_a P_a){\tilde Q}
  - g_{\rm{TC}}\,S\,\langle\bar{{\tilde Q}}{\tilde Q}\rangle  \nonumber \\
  && -\,\lambda_{\rm H}{\mathcal{H}}^4
  - \frac14\lambda_{\rm TC}(S^2+P^2)^2 +
  \lambda{\mathcal{H}}^2(S^2+P^2) \nonumber \\
  && +\, \frac12\mu^2_{\rm S}(S^2+P^2)+\mu_{\rm H}^2{\mathcal{H}}^2 \,,
  \label{LsM}
 \end{eqnarray}
with a particular choice of the ``source'' term linear in T-sigma 
where $\langle\bar{{\tilde Q}}{\tilde Q}\rangle<0$ is the diagonal T-quark condensate, 
${\mathcal{H}}^2=\mathcal{H}\mathcal{H}^\dagger$,
$P^2\equiv P_aP_a={\tilde \pi}^{0}{\tilde \pi}^{0}+2{\tilde \pi}^+{\tilde \pi}^-$, and the
EW-covariant derivatives are
 \begin{eqnarray} \nonumber
    &&\hat{D}{\tilde Q} = \gamma^{\mu} \left( \partial_{\mu}
       - \frac{iY_{{\tilde Q}}}{2}\, g'B_{\mu} - \frac{i}{2}\,
       g W_{\mu}^a \tau_a \right)\tilde{Q}\,, \\
    &&D_{\mu} P_a = \partial_{\mu} P_a + g
\epsilon_{abc} W^b_{\mu} P_c\,. \label{DQ}
 \end{eqnarray}
The Higgs boson doublet $\mathcal{H}$ in Eq.~(\ref{LsM}) acquires an 
interpretation as a composite bound state of vector-like T-quarks e.g. in the model 
extended by an extra $SU(2)_{\rm W}$-singlet Dirac ${\tilde S}$ T-quark such
that $\mathcal{H}={\tilde Q}\bar {\tilde S}$ (for other possibilities, see 
also Refs.~\cite{Cacciapaglia:2014uja,Hietanen:2014xca}). Note, however 
that at the moment the question about a particular UV content of the Higgs 
boson doublet is not of primary importance for the effective low-energy 
description of scalar T-meson interactions described 
by the phenomenological L$\sigma$M Lagrangian (\ref{LsM}) and thus 
will not be further discussed here.
\begin{figure*}[tbh]
\begin{minipage}{0.45\textwidth}
 \centerline{\includegraphics[width=1.0\textwidth]{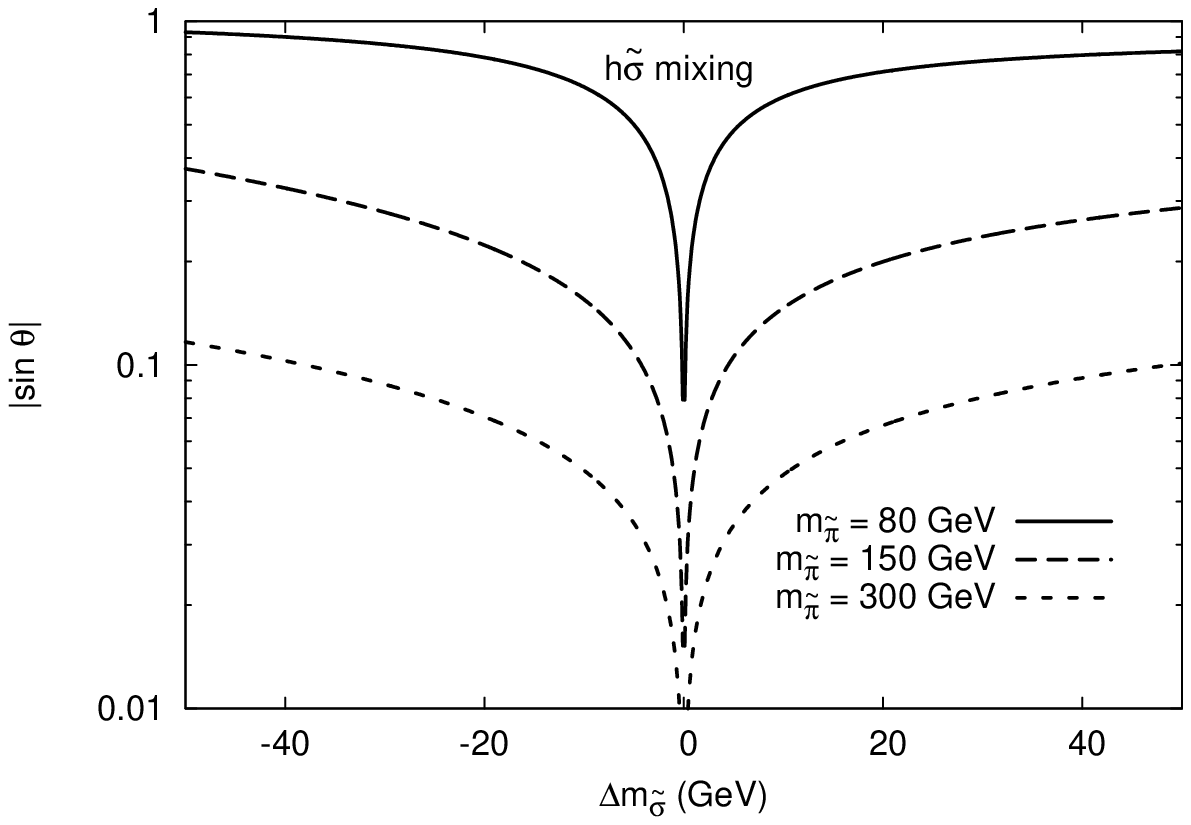}}
\end{minipage}
\hspace{1cm}
\begin{minipage}{0.45\textwidth}
 \centerline{\includegraphics[width=1.0\textwidth]{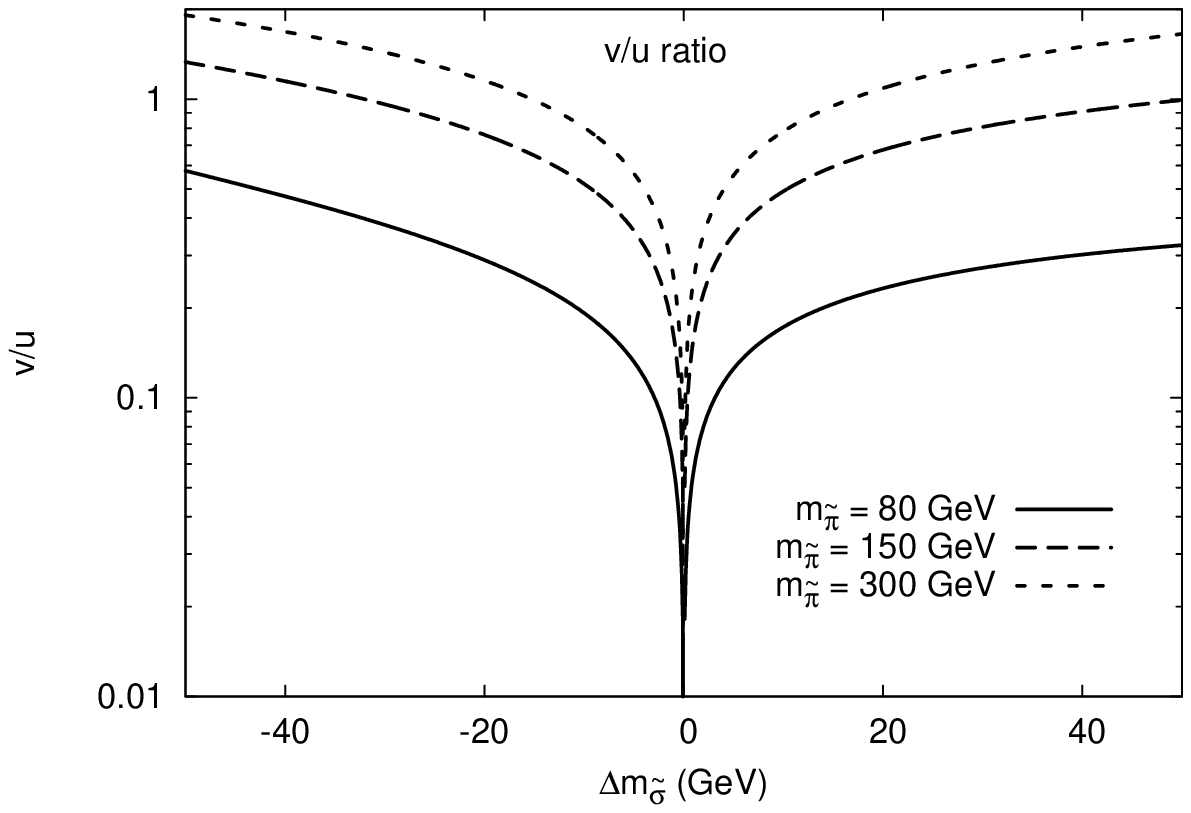}}
\end{minipage}
   \caption{The absolute value of sine of the $h\tilde{\sigma}$-mixing angle 
 $|\sin\theta|$ (left) and the ratio of the chiral and EW breaking scales $u/v$ (right)
 as functions of $\Delta m_{\tilde{\sigma}}\equiv m_{\tilde \sigma}-\sqrt{3}m_{\tilde{\pi}}$ 
 for three different values of the T-pion mass $m_{\tilde \pi}=80,150$ and 
 300 GeV. Here and below, the nearly-conformal limit has been imposed.}
 \label{fig:sin-th}
\end{figure*}

Note, the chiral symmetry implies the equality of constituent
Dirac masses $M_U=M_D\equiv M_{{\tilde Q}}$ at tree level. 
In the limit of small current T-quark masses $m_{\tilde Q}$ compared
to the constituent ones $M_{\tilde Q}$, i.e. $m_{\tilde Q}\ll M_{\tilde Q}\sim \Lambda_{\rm
TC}$, in analogy to ordinary QCD the conformal symmetry is
approximate such that the $\mu$-terms can be suppressed $\mu_{S,H}\ll
m_{{\tilde \pi}}$, which will be employed below throughout this work. 
Then the spontaneous EW and chiral symmetry breakings are initiated 
dynamically by the Higgs $v\simeq 246\,{\rm GeV}$ and T-sigma $u$ vevs 
\begin{eqnarray} \nonumber
 && \mathcal{H} = \frac{1}{\sqrt{2}}\left(
\begin{array}{c}
\sqrt{2}i\phi^-  \\
H+i\phi^0
\end{array}\right)\,,\quad \langle H\rangle \equiv v \,, 
\quad \langle S\rangle \equiv u \gtrsim v \,, \\
 && H=v+h c_\theta-\tilde \sigma s_\theta\,, \qquad  S = u + h s_\theta+\tilde \sigma c_\theta\,,
\label{shifts}
\end{eqnarray}
respectively, by means of T-quark condensation, namely,
 \begin{eqnarray} \nonumber
&&u=\left(\frac{g_{\rm{TC}}\lambda_{\rm H}}{\delta}\right)^{1/3}
|\langle\bar{{\tilde Q}}{\tilde Q}\rangle|^{1/3}\,, \\
&&v=\left(\frac{|\lambda|}{\lambda_{\rm H}}\right)^{1/2}
\left(\frac{g_{\rm{TC}}\lambda_{\rm H}}{\delta}\right)^{1/3}
|\langle\bar{{\tilde Q}}{\tilde Q}\rangle|^{1/3}\,, \label{u-v-min}
 \end{eqnarray}
and the T-pions acquire a mass
 \begin{eqnarray} \nonumber
m_{{\tilde \pi}}^2=-\frac{g_{\rm TC} \langle\bar{{\tilde Q}}{\tilde Q}\rangle}{u}\,.
 \end{eqnarray}
In the above expressions, $s_\theta\equiv \sin\theta$, $c_\theta\equiv \cos\theta$,
$\delta=\lambda_{\rm H}\lambda_{\rm{TC}} - \lambda^2$,
$g_{\rm{TC}}>0$ and $\lambda_H>0$. The minimal choice of the ``source'' 
term in the L$\sigma$M Lagrangian (\ref{LsM}) is natural since it simultaneously (i) sets up 
a pseudo-Goldstone mass scale for T-pions, (ii) allows to link all the incident vevs 
$u$ and $v$, and hence the constituent Dirac T-quark mass scale $M_{\tilde Q}=g_{\rm TC}u$, 
to the T-confinement scale, and (iii) describes the Yukawa interactions of the T-sigma with 
the diagonal T-quark condensate $\langle\bar{{\tilde Q}}{\tilde Q}\rangle$ which is the 
only dimensionfull nonperturbative parameter in the model at low energy scales.
The Nambu-Goldstone d.o.f.'s $\phi^{\pm},\phi^0$ originating from the Higgs doublet 
do not appear in the potential since ${\mathcal{H}}{\mathcal{H}}^\dagger=H^2/2$, 
where ${\mathcal{H}}$ and $H$ are defined in Eq.~(\ref{shifts}). They, thus, can not mix with pseudoscalar 
T-pions and get absorbed by the gauge bosons in normal way giving rise to longitudinal 
polarisations of $W^\pm,Z$ bosons, respectively. So, T-pions do not have tree-level 
couplings to SM fermions, and can only be produced in vector-boson fusion channels 
\cite{Pasechnik:2013bxa,Lebiedowicz:2013fta}.

In the VLC approach, one could distinguish two natural physical scales associated with
two chiral $u$ and EW $v$ symmetry breaking scales, which can, in principle, be very 
different from each other although are related to the single T-confinement scale, or 
the T-quark condensate $\langle\bar{{\tilde Q}}{\tilde Q}\rangle$.
In a phenomenologically natural hierarchy $u\gg v$, the upper scale would then roughly 
be associated with the mass scale of constituent T-quarks and T-baryons, whereas the 
lower one -- with the mass scale of lightest pseudo-Goldstone states, e.g. T-pions.
By a convention, in effective field theory approach one would then consider only 
the lightest states ($\tilde \pi$, $\tilde \sigma$) which propagate at short distances
and contribute to vacuum polarisations of the SM gauge bosons, while a net effect of all 
the heavy states (e.g. T-baryons, T-rho etc) can be effectively accounted for by constituent
T-quark loops. This is a natural consequence of imposing an upper cut-off in loop momentum 
scale $\mu\sim \Lambda_{\rm TC}\sim M_{\tilde Q}$ in the low-energy effective 
description suitable for e.g. oblique corrections calculation \cite{Pasechnik:2013bxa}.

In the framework of VLC model, the SM-like Higgs mechanism has an effective nature 
and is initiated by the Dirac T-quark condensation due to a presence 
of $\lambda{\mathcal{H}}^2S^2$ term in the potential (\ref{LsM}). 
While the Peskin-Tacheuchi $S$ and $U$ parameters \cite{Peskin-1,Peskin-2}
are strongly suppressed for all relevant model parameters $S,\,U \lesssim 0.01-0.001$, 
the T-sigma--Higgs mixing angle $\theta$ is bounded by the
$T$-parameter and the SM Higgs decay constraints provided that
$s_\theta\lesssim0.2$ \cite{Pasechnik:2013bxa}. In general, such a phenomenologically
consistent small $h\sigma$-mixing limit $s_\theta\to 0$ corresponds
to a decoupling of the TC dynamics from the SM up to higher energy
$\sim 1$ TeV scales, hence, to a suppressed ratio $v/u\ll 1$ as well as
to relatively weak TC couplings $g_{\rm TC},\lambda,\lambda_{\rm TC}\lesssim 1$ 
compared to analogical couplings in QCD. The latter property will be further
employed in the T-baryon sector.

In Fig.~\ref{fig:sin-th} the dependencies of sine of the $h\tilde{\sigma}$-mixing angle 
$|s_\theta|$ (left panel) and the Higgs and T-sigma vevs $v/u$ (right panel) on 
$\Delta m_{\tilde{\sigma}}\equiv m_{\tilde \sigma}-\sqrt{3}m_{\tilde{\pi}}$
are shown for three different T-pion mass values $m_{\tilde \pi}=80,150$ and 300 GeV.
Consequently, in the case of small $s_\theta\to 0$ and $v/u\ll 1$, or equivalently, 
$\Delta m_{\tilde{\sigma}}\to 0$, the deviations of the Higgs properties from those 
in the SM are small while the dynamical nature of the Higgs mechanism 
as a theoretically favorable possibility is preserved. The physical Lagrangian
of the VLC model can be found in Refs.~\cite{Pasechnik:2013bxa,Lebiedowicz:2013fta}
and we do not repeat it here.

In the currently favorable phenomenological situation with the SM-like Higgs boson, 
what would be the basic phenomenological signature for dynamical EW symmetry breaking? Besides the light 
Higgs boson, in the VLC model described above the T-pions are among 
the lightest physical T-hadron states which should be searched for in vector boson 
$VV$ and photon $\gamma\gamma$ fusion channels, preferably, in the low invariant 
mass region $m_{\tilde{\pi}}\sim 80 - 200$ GeV. The T-sigma state $\sigma$ is 
expected to be somewhat heavier since the small Higgs--T-sigma mixing limit 
$s_\theta\ll 1$ corresponds to $m_{\sigma}\sim \sqrt{3} m_{\tilde \pi}$ (Fig.~\ref{fig:sin-th}).
Besides, the T-sigma interactions with gauge bosons are strongly suppressed.
So, one of the most straightforward ways to search for the new strongly-coupled dynamics
and dynamical EW symmetry breaking in collider measurements is to look for 
T-pion signatures in $\gamma\gamma$-fusion channel \cite{Lebiedowicz:2013fta}.

Here, we discuss another source of constraints on such a new strong dynamics possibly
coming from astrophysics measurements at direct DM detection experiments. For this
purpose, let us consider the T-baryon spectrum of $SU(2)_{\rm TC}$ two-flavor theory.

\subsection{Scalar T-baryon Lagrangian}
\label{sec:TB-Lag}

Extra bound states of $SU(2)_{\rm TC}$ theory possessing an additional conserved (T-baryon) 
number analogous to the usual baryon number are given by scalar (anti)T-baryon multiplets $Q_iQ_j$ 
and ${\bar Q}_i{\bar Q}_j$. As was previously studied in Refs.~\cite{Lewis:2011zb,Appelquist:2014dja}, 
these states can play a role of pseudo-Goldstone bosons originating from global $SU(4)$ multiplets.
In the phenomenologically relevant TC decoupling limit $u\ll v$ the 
T-baryons can have a large mass $m_B\gtrsim M_{\tilde Q} \gg m_{\tilde \pi}, m_H$ 
significantly exceeding the EW breaking scale. In this case, just above the EW scale
one should employ the global chiral $SU(2)_L\otimes SU(2)_R$ symmetry in the T-meson 
sector whose effective Lagrangian is supplemented by an additional phenomenological 
SM-group invariant Lagrangian of heavy T-baryons. This case has not been discussed 
in the literature and deserves a special attention. And this is in variance with other 
somewhat similar UV completions not participating in the dynamical EW symmetry 
breaking such as those in Ref.~\cite{Buckley:2012ky}.

In order to describe consistently the EW and effective interactions of scalar T-baryons, 
let us start with two real adjoint (composite) representations of the SM $SU(2)_{\rm W}$ gauge
group $G_a$ and $F_a$, $a=1,2,3$, i.e. transforming in the weak
basis as
\begin{eqnarray}
G_a'=G_a-g\epsilon_{abc}\theta^b G_c\,, \quad
F_a'=F_a-g\epsilon_{abc}\theta^b F_c\,.
\end{eqnarray}
Introducing a complex adjoint representation $B$ with the unit
T-baryon charge and $Y_{\rm B}=0$ such as
\begin{eqnarray}
 B_a&=&\frac{1}{\sqrt{2}}(G_a+iF_a)\,,\quad B_a^*\equiv
\bar{B}_a\not=B_a\,, \\
 B_a'&=&B_a-g\epsilon_{abc}\theta^b B_c\,, \nonumber
\end{eqnarray}
one ends up with the gauge interactions of the scalar T-baryons 
$B_a$ set by the corresponding covariant derivative
\begin{eqnarray}
D_\mu B_a=\partial_\mu B_a+g\epsilon_{abc} W_\mu^b B_c\,.
\end{eqnarray}
In the charge basis, the physical T-baryon states are
\begin{eqnarray}
B^\pm=\frac{B_1\mp iB_2}{\sqrt{2}}\,, \;\; \bar B^\pm=\frac{\bar
B_1\pm i \bar B_2}{\sqrt{2}}\,, \;\; B^0=B_3\,.
\end{eqnarray}
Desirably, these states now have a definite partonic representation as diquark--like
bound states of $U,D$ T-quarks with conserved T-baryonic number
$T_B$, namely,
\begin{eqnarray*}
&& B^+=UU\,, \quad B^-=DD\,, \quad B^0=UD\,, \quad T_B=+1\,, \\
&& {\bar B}^+={\bar U}{\bar U}\,, \quad {\bar B}^-={\bar D}{\bar
D}\,, \quad {\bar B}^0={\bar U}{\bar D}\,, \quad T_{\bar B}=-1\,.
\end{eqnarray*}
For simplicity, at the first step in this work we assume that
energetically favored states for such composites are the ones with
$J=0$, although the lightest state with $J=1$ is not completely
excluded and will be studied elsewhere. The low-energy effective 
Lagrangian of T-baryon interactions with SM gauge bosons, 
T-sigma $S$ and T-pion $P^a,\,a=1,2,3$ invariant 
under local SM symmetry group additional to 
the L$\sigma$M Lagrangian (\ref{LsM}) reads
\begin{eqnarray} \nonumber
 \Delta{\cal L}_B &=& D_\mu\bar{B}_aD^\mu B_a +
 \frac12\, \mu_B^2 \bar B B +
 \frac12\, g_{\rm BB}(\bar B B)^2 \\ \nonumber
 &+& g_{\rm BS} (S^2+P^2)(\bar B B) +
 g_{\rm BH} (\mathcal{H}^\dagger \mathcal{H})(\bar B B) \\ &+&
 g_{\rm BP} (\bar B P) (B P)\,. \label{Delta-LB}
\end{eqnarray}
After the EW and CS breaking, the first four terms in the scalar
T-baryon potential give rise to the T-baryon mass term $m_{\rm
B}\bar BB$, where
\begin{eqnarray}
m_B^2=-2\mu_B^2 - 2g_{\rm BS} u^2 - g_{\rm BH} v^2 \,,
\end{eqnarray}
where, in general, $\mu_B^2<0$ and $g_{\rm BS},g_{\rm BH}$ couplings can be positive or negative. 
In the nearly-conformal limit $|\mu_B|\ll u,v$, one notices that the T-baryon mass
\begin{eqnarray}
m_B\gtrsim M_{\tilde Q}\sim u \,,
\end{eqnarray}
is naturally large in the TC decoupling limit $u\gg v$ and $g_{\rm BS}<0$.
In practice, one arrives at the model with five physical parameters 
controlling the properties of scalar T-baryons: the T-baryon mass 
scale $m_B\gtrsim \Lambda_{\rm TC}\sim 1$ TeV and four 
T-baryon--scalar dimensionless couplings $g_{{\rm
B},i} \equiv \{g_{\rm BB},\,g_{\rm BH},\,g_{\rm BS},\,g_{\rm BP}\}$. Remind, the masses of 
$\tilde\pi$, $\tilde \sigma$ and $h$ bosons remain light and are placed 
at the EW scale in the heavy T-baryon limit under discussion \cite{Pasechnik:2013bxa},
thus, validating the suggested scenario.
\begin{figure*}[tbh]
 \centerline{\includegraphics[width=0.8\textwidth]{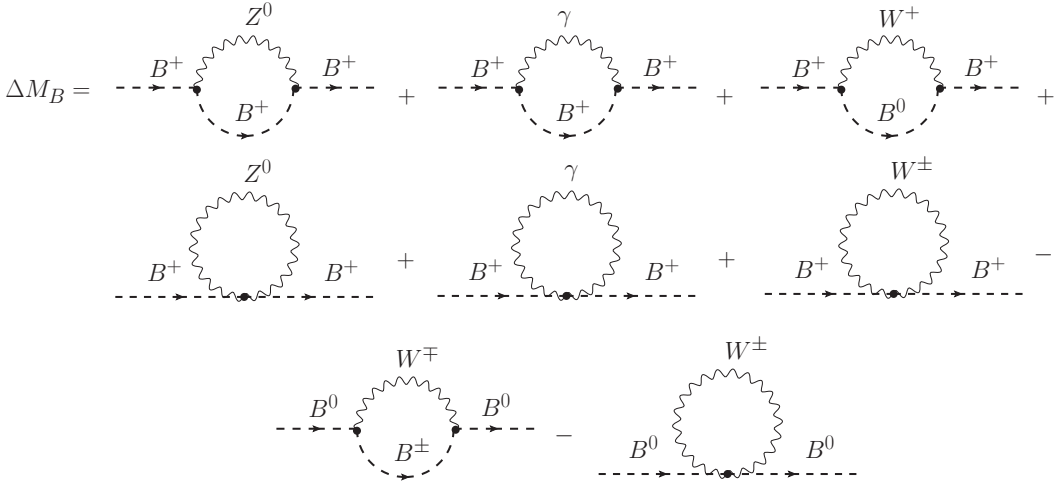}}
   \caption{
 \small Diagrams contributing to the EW mass splitting $\Delta m_B$ 
 in the scalar T-baryon spectrum.}
 \label{fig:mass-split}
\end{figure*}

The meson-meson and baryon-meson interactions in ordinary QCD are usually 
very strong and non-perturbative i.e. the corresponding couplings are typically 
much larger than unity. Naively, this happens since the confinement scale is 
turned out to be comparable to the mass scale of pseudo-Goldstone modes 
in QCD. One sometimes refers to a very dense pion cloud surrounding nucleons 
as to a cause for such large nonperturbative couplings. In the case of relatively
light (possibly, composite) Higgs boson and T-pions $m_{\tilde \pi}\sim 100-150$
GeV compared to a large T-confinement scale $\Lambda_{\rm TC}\gtrsim
1$ TeV, it may be more natural to expect a very different situation in
T-hadron interactions -- the T-pion cloud surrounding T-baryons is likely
to be rather loose or sparse for $v/u\ll 1$ and $s_\theta\ll 1$. Thus, the T-hadron
interactions at low energies may not be as intense as in QCD and the
corresponding couplings should be smaller in this case although a more 
dedicated (e.g. lattice) analysis, of course, is necessary. The latter 
argument is in full agreement with the consistent small $h\tilde \sigma$-mixing limit 
mentioned above where the scalar self-couplings turn out to be small or even vanishing, 
and thus one expects the T-baryon--T-meson couplings to be small as well, i.e. $g_{{\rm
B},i}\lesssim 1$.

The low-energy effective Lagrangian (\ref{Delta-LB}) appears to have an
extra exact global $U(1)_{\rm TB}$ symmetry corresponding to the
T-baryon number conservation in analogy to ordinary baryon symmetry
in the SM. Note, this symmetry has not been imposed forcefully.
Instead, it emerged automatically due to the structure of T-baryon
multiplet in the complex $SU(2)$-adjoint representation imposed by
the chosen Dirac UV completion. In analogy to the usual baryon
abundance, however, the present DM abundance may originate via
generation of a sufficient T-baryon asymmetry e.g. by means of the
well-known EW sphaleron mechanism at high energies in early
Universe, in accordance to the popular ADM scenario 
\cite{Lewis:2011zb,Belyaev:2010kp,Petraki:2013wwa}.
\begin{figure*}[tbh]
 \centerline{\includegraphics[width=0.9\textwidth]{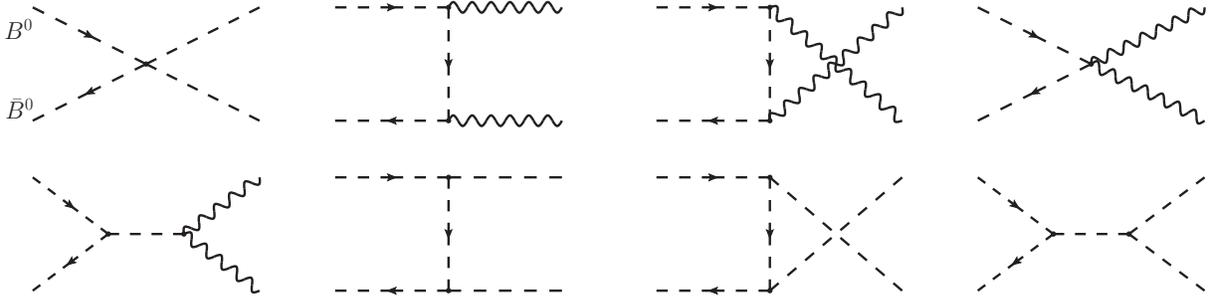}}
   \caption{
 \small Typical topologies for scalar T-baryon annihilation in the
 high-symmetry (HS) phase, $B^0\bar B^0\to W^aW^b,{\cal H}{\cal H},\,SS,\,P_aP_b$, $a,b=1,2,3$ (upper row), 
 and low-symmetry (LS) phase, $B^0\bar B^0\to W^+W^-,\,S_1S_2$, where $S_{1,2}=\tilde \pi^{\pm,0},\tilde \sigma,h$ 
 (both upper and lower rows).}
 \label{fig:TB-annihilation}
\end{figure*}

The physical Lagrangian derived from Eq.~(\ref{Delta-LB}) describes the T-baryon interactions
with gauge $W^\pm,Z,\gamma$ bosons, scalar Higgs $h$ and T-sigma
$\tilde \sigma$ and T-pions ${\tilde \pi}^0,{\tilde \pi}^\pm$,
as well as their self-interactions. The full expression is rather
lengthy, and here we show the terms most relevant for T-baryon
phenomenology only. In particular, the parameter-free gauge interactions
critically important for possible T-baryon production mechanisms at
the LHC and T-baryon scattering off nucleons are given by
\begin{eqnarray}
 && {\cal L}_{\rm VB\bar B} = ig\Big[W_\mu^-(B_{,\mu}^0 \bar B^- -
B^0\bar B_{,\mu}^- + \bar B_{,\mu}^0 B^+ - \bar B^0B_{,\mu}^+)  \nonumber \\
&& +\, (s_W A_\mu + \, c_W Z_\mu)(B_{,\mu}^+ \bar B^+ + \bar
B_{,\mu}^- B^-)\Big] + c.c. \,, \label{VBB} \\
 && {\cal L}_{\rm VVB\bar B} =
g^2\Big[(\bar B^0B^0+\bar B^+ B^+) \, W_\mu^+{W^\mu}^- \nonumber \\
&& +\, \bar B^+ B^+
(s_WA_\mu + c_WZ_\mu)^2 - \bar B^+B^-W_\mu^+{W^\mu}^+ \label{VVBB}  \\
&& -\, (\bar B^+B^0 + \bar B^0 B^-)(s_WA_\mu +
c_WZ_\mu){W^\mu}^+\Big] + c.c.\,, \nonumber
\end{eqnarray}
with the SM parameters $g$ and $\theta_W$ fixed at the T-baryon mass
scale $\mu=m_{\rm B}$ as a low-energy limit of the T-baryonic gauge form
factors. Note, vector $\bar B^0 B^0 Z$-type couplings which determine the
SI scattering of DM particles off nucleons at the Born level do not
exist in this scenario. The scalar T-baryon--quark scattering, in fact, 
enters via an exchange by $W$-boson pair at one-loop level 
while the T-baryon--gluon scattering -- at two-loop level with 
an intermediate charged T-baryon, and thus are strongly suppressed. 
The scalar T-baryon DM in this sense has a certain similarity to the 
electroweak-interacting DM scenario discussed e.g. in Ref.~\cite{Hisano:2011cs}.
\begin{figure*}[tbh]
\begin{minipage}{0.45\textwidth}
 \centerline{\includegraphics[width=1.0\textwidth]{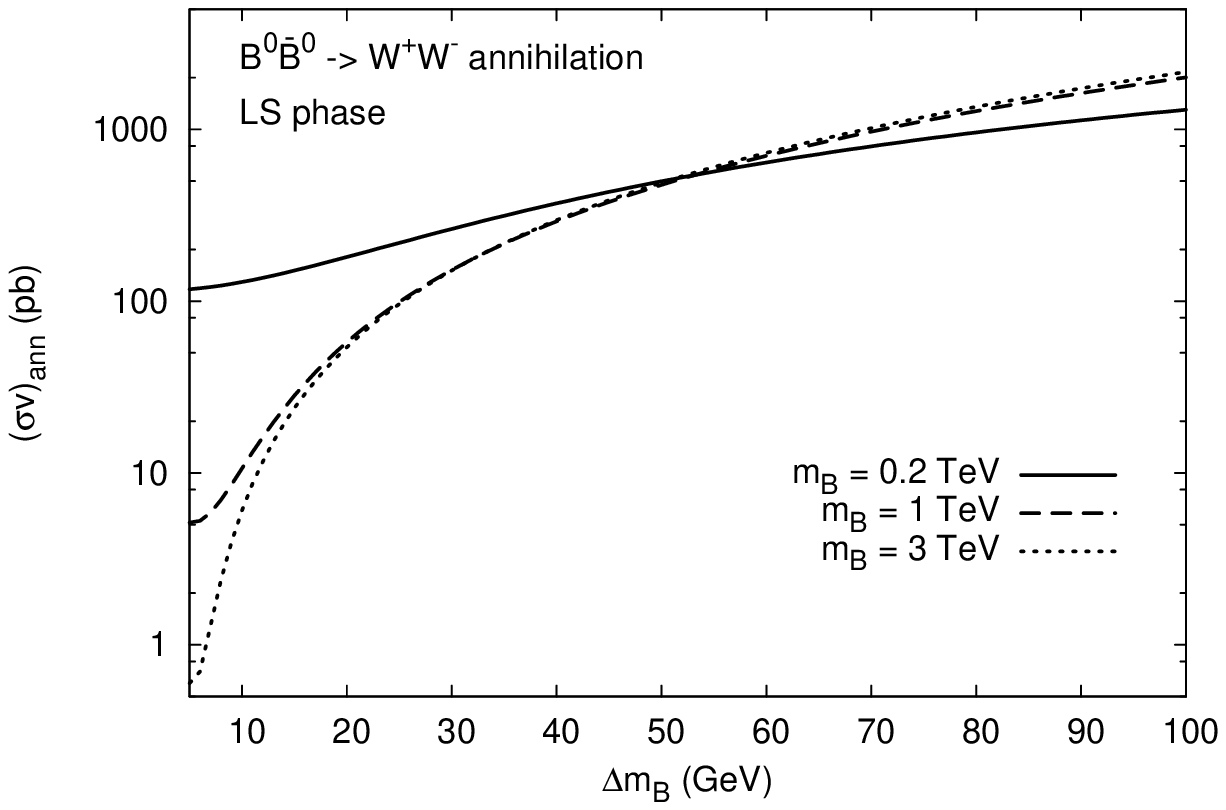}}
\end{minipage}
\hspace{1cm}
\begin{minipage}{0.45\textwidth}
 \centerline{\includegraphics[width=1.0\textwidth]{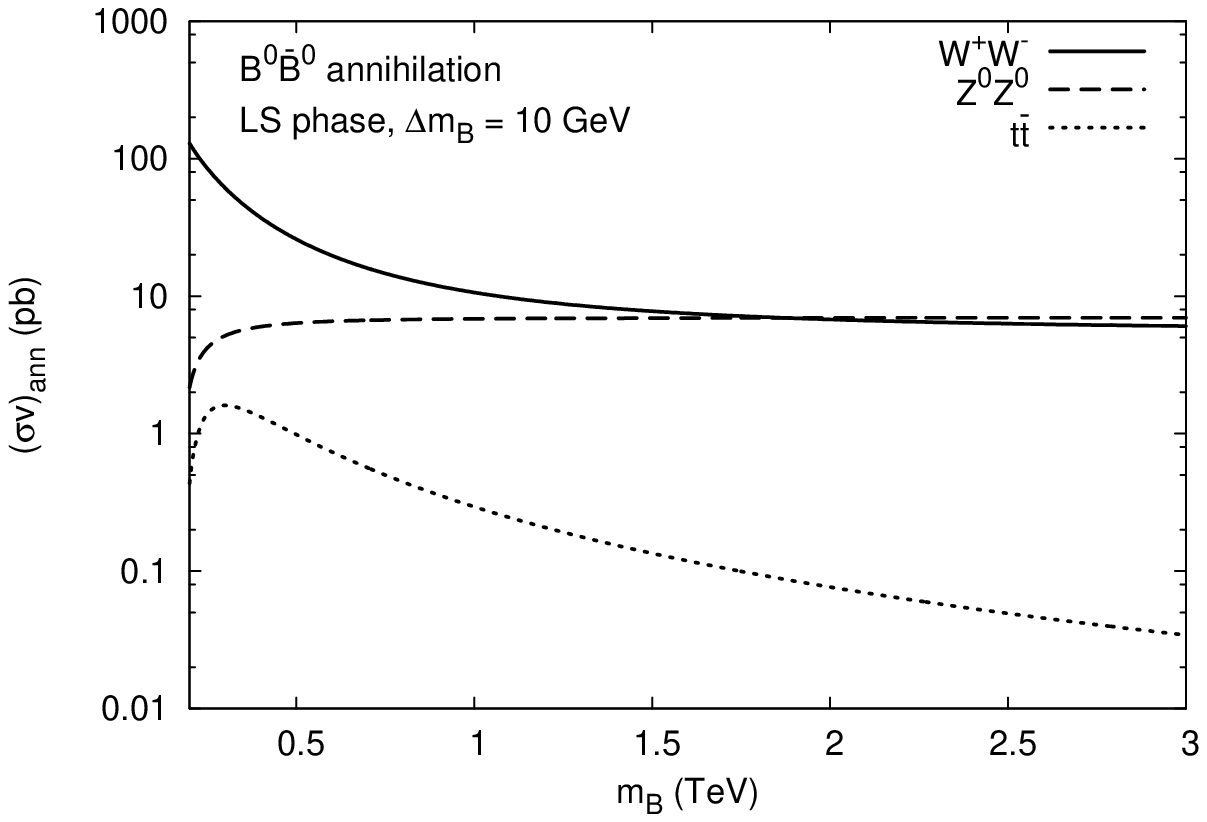}}
\end{minipage}
   \caption{The T-baryon LS annihilation cross section in the $B^0\bar B^0\to W^+W^-$ 
   channel as a function of the T-baryon mass splitting $\Delta m_B$ (left) and the largest
   annihilation channels into SM final states $B^0\bar B^0\to W^+W^-,Z^0Z^0,t\bar t$ 
   as functions of the T-baryon mass scale $m_B$ (right). In the left panel, the results
   are shown for three different mass scales $m_B=0.2,1,3$ TeV, in the right panel 
   the $W^+W^-$ contribution is shown for $\Delta m_B=10$ GeV. Here, 
   $m_{\tilde{\pi}}=150$ GeV and $\Delta m_{\tilde{\sigma}}=20$ GeV are fixed.}
 \label{fig:SM-ann}
\end{figure*}

The scalar T-baryon annihilation processes in early Universe are
largely dominated by effective T-baryon interactions with light
composites ($\tilde \sigma$ and $\tilde \pi$) and the Higgs boson given by
\begin{eqnarray} \nonumber
{\cal L}_{\rm BS}&=&\bar B B \big\{ g_{\rm BS}\big[S_1^2 + 2uS_1
+ P^2 \big] + \frac12\, g_{\rm BH}\big[S_2^2 + 2vS_2\big] \big\} \\
&+& g_{\rm BP}\big[\bar{B}^0B^0{\tilde \pi}^0{\tilde \pi}^0 + (\bar{B}^+B^+ +
\bar{B}^-B^-){\tilde \pi}^+{\tilde \pi}^- \nonumber \\ &+& (\bar{B}^-{\tilde \pi}^- +
\bar{B}^+{\tilde \pi}^+)B^0{\tilde \pi}^0 + (B^+{\tilde \pi}^- + B^-{\tilde \pi}^+)\bar B^0{\tilde \pi}^0
\nonumber \\ &+& \bar{B}^-B^+{\tilde \pi}^-{\tilde \pi}^- + \bar{B}^+B^-{\tilde \pi}^+{\tilde \pi}^+
\big] \,,
\end{eqnarray}
where $S_1 = h s_\theta + \tilde \sigma c_\theta$ and $S_2 = h c_\theta -
\tilde \sigma s_\theta$. The latter Lagrangian is thus critical for the Higgs-induced
tree-level elastic scattering of scalar T-baryons off nucleons in direct DM 
detection measurements as well as for the formation of 
the relic T-baryon abundance.

\section{Scalar T-baryon Dark Matter}

\subsection{T-baryon mass splitting}

As was mentioned above, the T-baryon states in the strongly coupled
$SU(2)_{\rm TC}$ theory are scalar composite di-T-quark states and
can be rather heavy in the TC decoupling limit, $u\gg v$.

At temperatures above the EW phase transition scale but below
the chiral symmetry breaking scale, $u\gg T\gtrsim v$, the $UD$ state remains
degenerate with other components of the $SU(2)$ triplet $B^a$. This is valid
in the framework of considering low-energy effective theory where
one-loop self-energies with virtual T-baryons, T-pions, Higgs boson and 
T-sigma are the same for $B^0$ and $B^\pm$ and thus do not contribute 
to the mass splitting between neutral and charged T-baryons. 

At lower temperatures, $T < v$, corresponding to the the EW-broken phase 
the di-T-quark state with zeroth electric charge $UD$ is expected to be energetically 
favorable. Together with exact global T-baryon symmetry, the latter 
suggests that the neutral di-T-quark $B^0\equiv UD$ state 
is also the lightest and thus stable. The mass splitting between $B^0$ and $B^\pm$
at low temperatures get well-defined contribution from the EW corrections depicted 
in Fig.~\ref{fig:mass-split}. The latter are given by the one- and two-point self-energies 
with gauge bosons in the loop which are not cancelled in the mass 
difference according to Eqs.~(\ref{VBB}) and (\ref{VVBB}).
The result for EW-induced mass splitting reads
\begin{widetext}
\begin{eqnarray} \nonumber
\Delta m^{\rm EW}_B & = & \frac{g_2^2 m_W^2}{16\pi^2 m_B} 
\Big\{ \ln\Big(\frac{m_Z^2}{m_W^2}\Big) - \beta^2(\mu_Z) \ln(\mu_Z) + \beta^2(\mu_W) \ln(\mu_W) \\
&-& \frac{4\beta^3(\mu_Z)}{\sqrt{\mu_Z}}\, \Big[\arctan\Big(\frac{\sqrt{\mu_Z}}{2\beta(\mu_Z)}\Big) + 
    \arctan\Big(\frac{2-\mu_Z}{2\sqrt{\mu_Z}\beta(\mu_Z)}\Big)  \Big] \nonumber \\
&+& \frac{4\beta^3(\mu_W)}{\sqrt{\mu_W}}\, \Big[\arctan\Big(\frac{\sqrt{\mu_W}}{2\beta(\mu_W)}\Big) +
    \arctan\Big(\frac{2-\mu_W}{2\sqrt{\mu_W}\beta(\mu_W)}\Big) \Big] \Big\} \,,
\end{eqnarray}
\end{widetext}
where $\mu_{Z/W}=m_{Z/W}^2/m_B^2$ and $\beta(x)=\sqrt{1-x^2/4}$. 
In the realistic limits, $\Delta m_B\ll m_B$ and $m_{Z/W}\ll m_B$, the EW mass splitting
may be estimated as
\begin{eqnarray}
\Delta m^{\rm EW}_B \simeq \frac{g_2^2}{8\pi}\, m_W (1-c_W) \approx 0.17\, \mathrm{GeV}\,. \label{EW-split}
\end{eqnarray}

Note, at the perturbative level the T-strong-induced mass splitting vanishes 
in both EW-unbroken and EW-broken phases in the considering model. However, 
in analogy to the di-quark spectrum QCD, one may discuss potentially large non-perturbative 
T-strong effects in the mass splitting in the chiral/EW symmetry broken phase. 
The situation is close to what we have in ordinary QCD when $ud$ di-quark 
is split down in the di-quark mass spectrum by as much as $\sim 70$ MeV 
due to pion exchanges. Such non-perturbative effects, should be a proper subject 
for lattice studies and are not discussed here. In any case, the EW-induced 
splitting above (\ref{EW-split}) can be safely treated as a conservative lower bound. 
So for the sake of generality in our numerical analysis we consider the T-baryon DM implications 
in the EW-broken phase over the following wide range of allowed mass splittings:
\[ \Delta m^{\rm EW}_B < \Delta m_B < M_{\rm EW} \sim  100\, {\rm GeV} \,, \]
while the basic qualitative conclusions will not strongly depend on the actual 
value of $\Delta m_B$.

\subsection{Cosmological evolution}

First, consider cosmological evolution of the T-baryon density in
early Universe determined by the T-baryon annihilation processes.
An assumption that the DM today consists mainly of scalar T-baryons 
provides with the upper limit on relic T-baryon abundance \cite{Steigman:2012nb}
\begin{eqnarray}
 && \Omega_{\rm TB}\lesssim \Omega_{\rm CDM}\simeq 0.2\Big[(\sigma
v)^{\rm DM}_{\rm ann}/(\sigma v)^{\rm th}_{\rm ann}\Big]\,, \nonumber \\
 &&(\sigma v)_{\rm ann}^{\rm DM}\simeq 2.0\times 10^{-9}\; {\rm
GeV}^{-2}\simeq 0.78\, {\rm pb}\,. \label{omega-TB}
\end{eqnarray}
The combined data from WMAP nine-year mission \cite{Hinshaw:2012aka} suggest 
the value for present-day relic DM abundance $\Omega_{\rm CDM}h^2=0.1138\pm 0.0045$.
The expression (\ref{omega-TB}) is then considered as a source for canonical constraints on 
TC model parameters and T-baryon asymmetry as long as a theoretical prediction for thermally
averaged kinetic T-baryon annihilation cross section $(\sigma
v)_{\rm ann}^{\rm th}$ is provided.

The irreversible T-baryon annihilation effectively starts at temperature $T_i\simeq m_B/3$.
Depending on $T_i$, as well as on the T-baryon freeze-out temperature, $T_f\simeq m_B/20$, 
compared to the EW phase transition temperature, $T_{\rm EW}\sim M_{\rm EW} \sim 100$ GeV 
one may consider two limiting possibilities -- the annihilation of T-baryons in the high
($T_f\gtrsim T_{\rm EW}$) and low ($T_i\lesssim T_{\rm EW}$) symmetry phases of the
cosmological plasma (for more details on the characteristics of DM annihilation 
in these phases, see Ref.~\cite{Pasechnik:2013kya}). 
\begin{figure}[h!]
 \centerline{\includegraphics[width=0.5\textwidth]{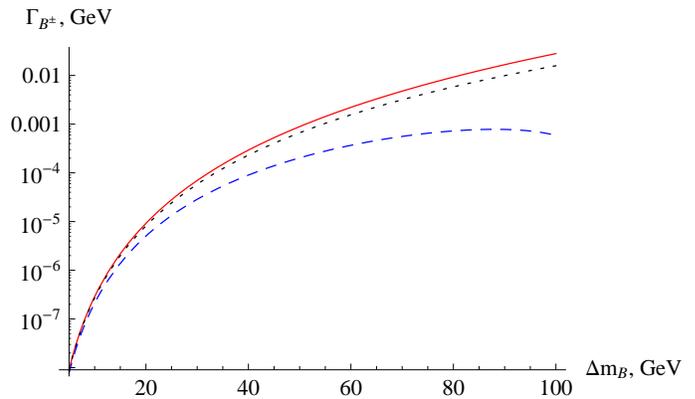}}
   \caption{
 \small Total decay width of charge T-baryons $B^\pm \to B^0 + 
 W^\pm(\to l \bar \nu_l,\,q_i \bar q_j)$ as a function of the mass 
 splitting $\Delta m_B$ for different T-baryon mass scales $m_B=100$ GeV 
 (dashed), $m_B=500$ GeV (dotted) and $m_B=10$ TeV (solid).}
 \label{fig:TB-decay}
\end{figure}
\begin{figure*}[tbh]
\begin{minipage}{0.45\textwidth}
 \centerline{\includegraphics[width=1.0\textwidth]{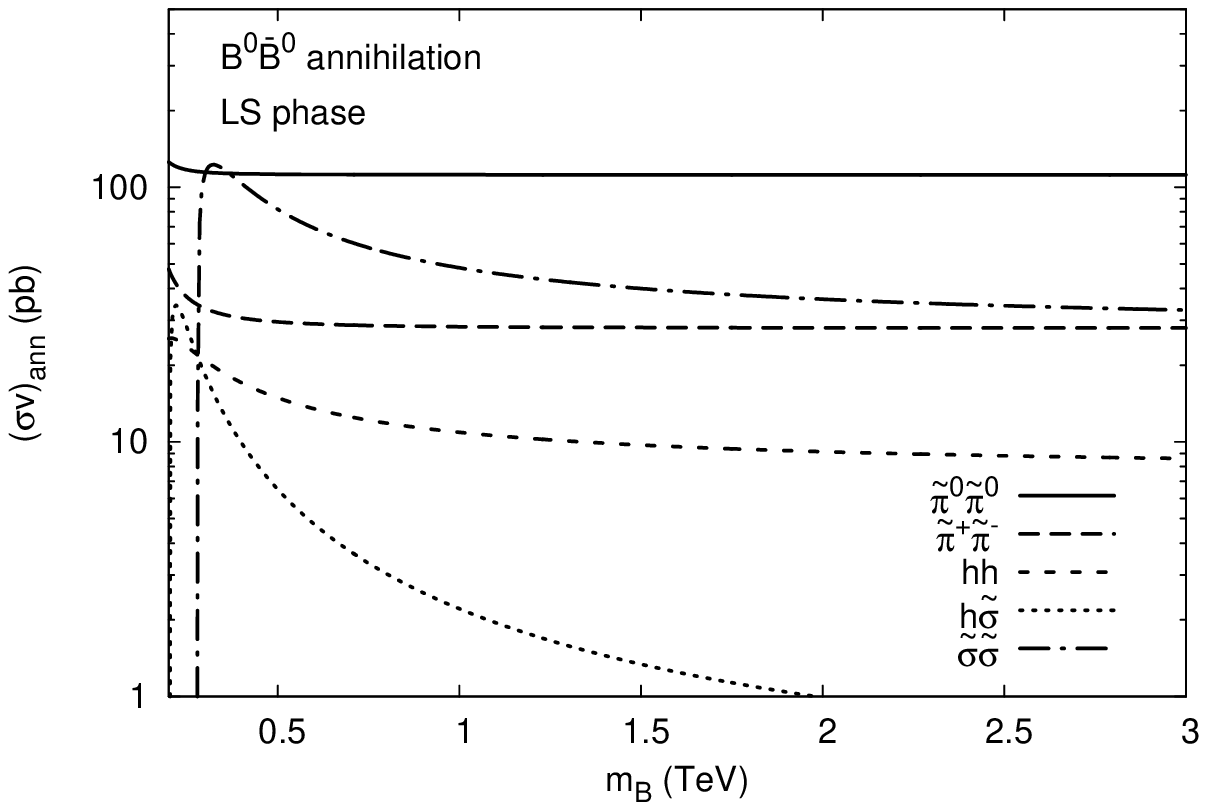}}
\end{minipage}
\hspace{1cm}
\begin{minipage}{0.45\textwidth}
 \centerline{\includegraphics[width=1.0\textwidth]{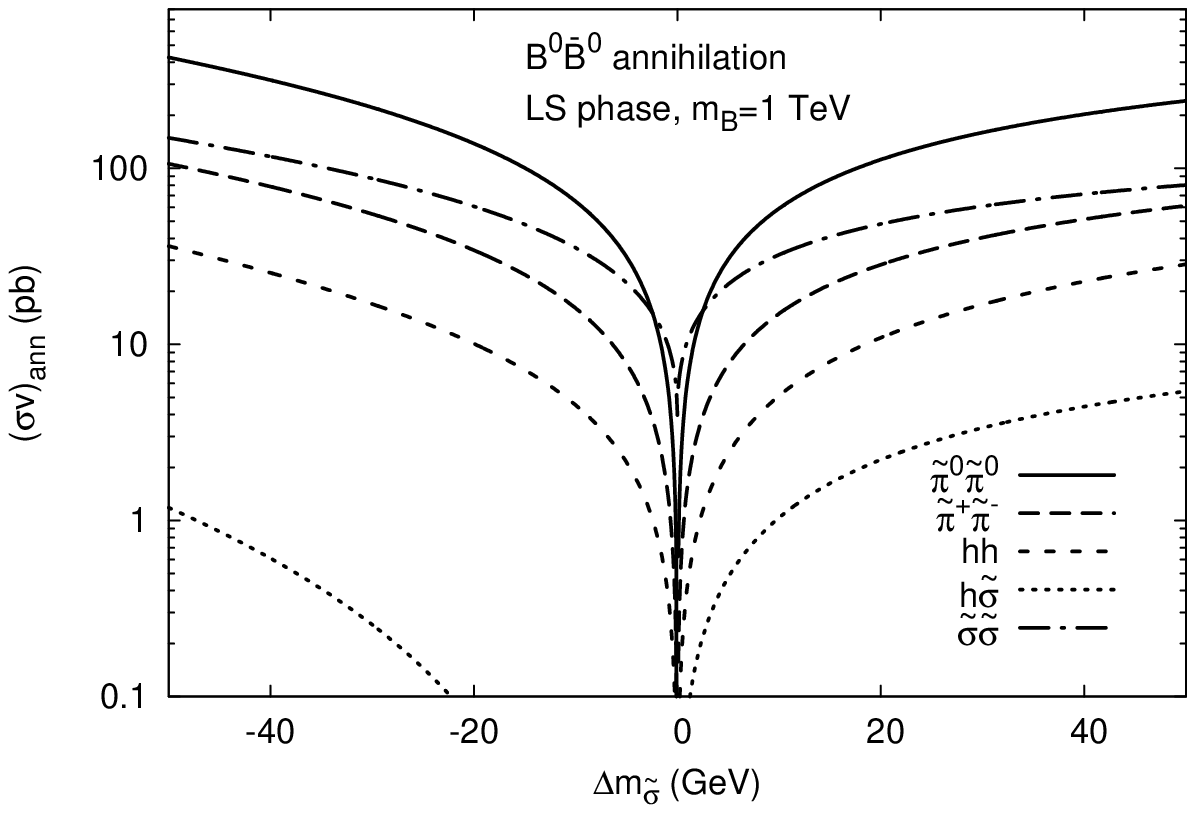}}
\end{minipage}
   \caption{The T-baryon LS annihilation cross sections in the (pseudo)scalar $B^0\bar B^0\to 
   \tilde{\pi}^0\tilde{\pi}^0,\tilde{\pi}^-\tilde{\pi}^-,hh,h\tilde{\sigma},\tilde{\sigma}\tilde{\sigma}$ 
   channels as functions of the T-baryon mass splitting $m_B$ for fixed $\Delta m_{\tilde{\sigma}}=20$ 
   GeV (left) and $\Delta m_{\tilde{\sigma}}\equiv m_{\tilde \sigma}-\sqrt{3}m_{\tilde{\pi}}$ for 
   fixed $m_B=1$ TeV (right). Here, $m_{\tilde{\pi}}=150$ GeV is fixed.}
 \label{fig:TC-ann-mB_DMS}
\end{figure*}

Below the EWSB temperature, all heavier charged T-baryon $B^\pm$ and $\bar B^\pm$ states
are expected to decay very quickly $B^\pm \to B^0 + W^\pm(\to l \bar \nu_l,\, 
q_i \bar q_j)$, and thus do not participate in annihilation processes.
For example, the total leptonic $B^\pm$ 3-body decay width reads
\begin{eqnarray} \nonumber
\Gamma^{lept}_{B^\pm}&=&\sum_{l=e,\mu}\frac{g_2^4}{768\pi^3m_W^4m_\pm}
\int^{(\Delta m_B)^2}_{m_l^2}\Big(1-\frac{m_l^2}{q^2}\Big)
\\ &\times& \bar{\lambda}(q^2,m_B^2,m_\pm^2)F(q^2)dq^2\,,
\end{eqnarray}
where $\tau\nu_\tau$ channel does not contribute since $\Delta m_B< m_\tau$, and
\begin{eqnarray} \nonumber
F(q^2)&=& (m_\pm^2-m_B^2)^2\Big(1+\frac{m_l^2}{q^2}-
2\frac{m_l^4}{q^4}\Big) \\
&-& q^2\big[2(m_\pm^2+m_B^2)-q^2\big]
\Big(1-\frac{m_l^2}{2q^2}-\frac{m_l^4}{2q^4}\Big) \,, \nonumber \\
\bar{\lambda}(a,b,c)&=& \Big(1-2\frac{a+b}{c}+
\frac{(a-b)^2}{c^2}\Big)^{1/2}\,, \label{lambda}
\end{eqnarray}
and similarly for quark channels (additionally multiplied by the corresponding 
CKM matrix elements squared $|V_{ij}|^2$). 

In the case of small mass splitting (\ref{EW-split}), the total $B^\pm\to B^0+f_i\bar f_j$ decay width 
into light fermions is not practically sensitive to the T-baryon mass $m_B$, 
and is approximately equal to
\[
\Gamma_{B^\pm}\simeq 1.7\times 10^{-16}\, \mathrm{GeV} \,, \qquad \Delta m_B = \Delta m^{\rm EW}_B \,.
\]
which corresponds to the $B^\pm$ mean lifetime in the cosmological plasma,
$\tau_B\simeq 3.8 \times 10^{-9}$ s. This lifetime is larger 
than the Hubble time at the DM freeze-out,
$\tau_f\equiv H^{-1}(T_f)\lesssim 10^{-10}$ s for $m_B\gtrsim 1$ TeV.
This means that both neutral $B^0$ and $B^\pm$ components
participate in the DM annihilation and such co-annihilation reactions as 
$B^0\bar B^\pm \to X$, $B^\mp\bar B^\pm \to X$ etc contribute to the formation 
of DM relic density. Given that the $SU(2)$ symmetry is nearly restored in the T-baryon sector
in the limit $\Delta m_B \ll M_{\rm EW}$ and the decoupling limit $M_{\rm EW}\ll \Lambda_{\rm TC}$ 
is concerned, to a good approximation one could safely employ the T-baryon annihilation cross 
section $B^a\bar B^a\to X$ obtained in the EW unbroken phase (see below).

In the case of large mass splitting $\Delta m_B \gg m_{l,q}$, the partial $B^\pm$ decay widths into light fermions 
are not sensitive to the fermion masses $m_{l,q}$, and can be accounted for by a multiplicative
factor in the total decay width $\Gamma_{B^\pm}$. The latter is shown in Fig.~\ref{fig:TB-decay} as 
a function of $\Delta m_B=[10\dots 100]$ GeV for three distinct T-baryon mass scales $m_B=0.2,\,1,\,10$ TeV
by dashed, dash-dotted and solid lines, respectively. There is a rather strong dependence of 
the decay rate on the T-baryon mass splitting, whereas almost no sensitivity is found w.r.t. 
the mass scale $m_B$ variations. For TeV-mass T-baryons, it is straightforward 
to estimate the realistic mean lifetime of charge $B^\pm$ in the cosmological plasma 
$\tau_B=\Gamma_{B^\pm}^{-1}\sim 10^{-23}-10^{-22}$ s for a realistic splitting $\Delta m_B\sim 100$ GeV 
which is about ten orders of magnitude smaller than the Hubble time in the beginning of DM annihilation 
epoch $H^{-1}(T_i)\sim 10^{-12}$ s. This means that at temperatures below $T_i$ the charge 
T-baryons do not present in the plasma and, thus, no co-annihilation reactions 
$B^0\bar B^\pm \to X$, $B^\mp\bar B^\pm \to X$ etc contribute to the formation 
of DM relic density. So, only $B^0\bar B^0\to X$ process should be considered which
simplifies the subsequent calculations.

In analogy to ordinary baryons, the existence of a sufficient initial T-baryon asymmetry can 
be critical for a non-negligible amount of the scalar T-baryon DM in the case of large 
T-baryon--scalar couplings. So the T-baryon asymmetry appears to be an important 
parameter which completely determines the amount of such DM in the present 
Universe in the case of too fast annihilation rates. Various types of 
mixed (partially ADM) scenarios refer to an intermediate
case of noticeable amounts of both particles and antiparticles although in not exactly equal
amounts. A large existing freedom in the choice of T-baryon--scalar couplings 
$g_{{\rm B},i}$ makes it possible to consider both ADM and mixed DM scenarios.
Below, for simplicity in our numerical analysis we consider a naive VLC scenario 
(in the near-conformal regime) where all the scalar self-couplings are equal to $g_{\rm TC}$, i.e.
\begin{eqnarray} \label{gB-coupl}
g_{{\rm B},i}=g_{\rm TC}=\frac{M_{\tilde{Q}}}{u}\,, \quad M_{\tilde{Q}}\simeq \frac{m_B}{2}\,,
\end{eqnarray}
which contains basics features of the general case. In this simplified scenario, 
$m_B$, $m_{\tilde \pi}$, and $\Delta m_{\tilde{\sigma}}\equiv
m_{\tilde \sigma}-\sqrt{3}m_{\tilde{\pi}}$ are the only variable 
independent parameters.

Dominating topologies for EW and T-strong annihilation channels are
shown in Fig.~\ref{fig:TB-annihilation}. In the {\it high-symmetry phase} (or HS phase) valid for
heavy T-baryons $m_B\gtrsim 2$ TeV, the weak-induced T-baryon annihilation 
goes through the process $B^0\bar B^0\to W_1W_1+W_2W_2$ 
(with massless $W^a$) whose cross section reads
\begin{eqnarray} \label{HS-W}
\sigma^{\rm W,HS}_{\rm ann} &=& \frac{2g_2^4}{\pi s \sqrt{1-\frac{4m_B^2}{s}}}
\Big[\Big(5+\frac{4m_B^2}{s} \Big)\sqrt{1-\frac{4m_B^2}{s}} \\ &+& 
\Big(1-\frac{6m_B^2}{s} + \frac{64m_B^4}{s^2} \Big)\ln 
\frac{1+\sqrt{1-\frac{4m_B^2}{s}}}{1-\sqrt{1-\frac{4m_B^2}{s}}}\Big] \,.
\nonumber
\end{eqnarray}
In the non-relativistic limit $v\ll 1$, however, the corresponding kinetic cross section
\begin{eqnarray} \label{HS-W-v}
(\sigma v)^{\rm W,HS}_{\rm ann} \simeq \frac{13 g_2^4}{2\pi m_B^2}\, v
\end{eqnarray}
vanishes. So, in this case the total annihilation cross section $(\sigma v)^{\rm th}_{\rm ann}$ 
which enter Eq.~(\ref{omega-TB}) will be essentially determined by 
the TC-induced $B^0\bar B^0\to {\cal H}{\cal H},\,SS,\,P_aP_a$ 
channels (with massless final states), i.e.
\begin{eqnarray} \label{HS-TC}
(\sigma v)^{\rm th}_{\rm ann}\simeq (\sigma v)^{\rm TC,HS}_{\rm ann} &\simeq& 
\frac{g_{\rm HS}^2}{16\pi m_B^2} \,,
\end{eqnarray}
where
\begin{eqnarray} \nonumber
g_{\rm HS}^2 &=& \frac14 g_{\rm BH}^2 + 
3 g_{\rm BS}^2 + (g_{\rm BS} + g_{\rm BP})^2\,.
\end{eqnarray}
Here, the diagrams with a single 4-particle vertex dominate, others are strongly
suppressed by extra powers of $m_B$ in propagators. Note, an inclusion 
of extra possible composites to the theory, e.g. in extended chiral symmetries 
and composite Higgs scenarios, as well as (pseudo)vector T-mesons may
only increase the annihilation cross section. 
\begin{figure*}[tbh]
\begin{minipage}{0.45\textwidth}
 \centerline{\includegraphics[width=1.0\textwidth]{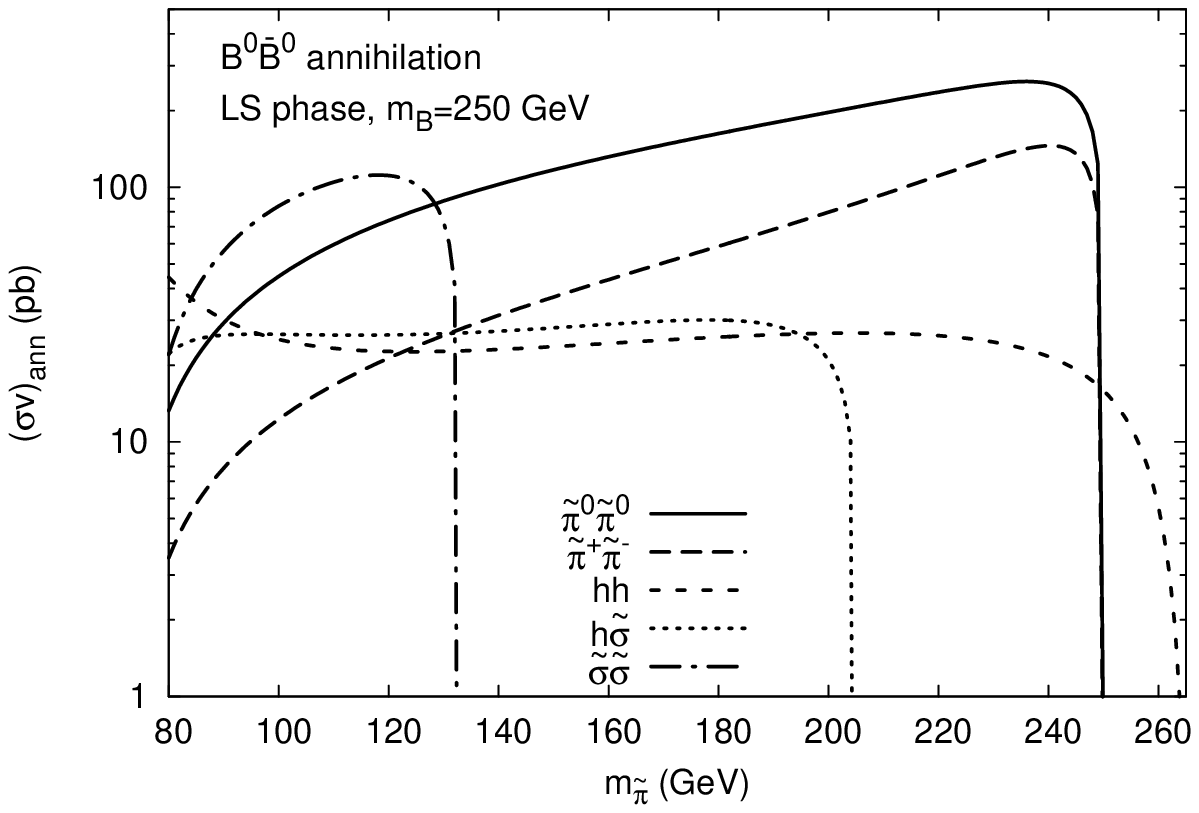}}
\end{minipage}
\hspace{1cm}
\begin{minipage}{0.45\textwidth}
 \centerline{\includegraphics[width=1.0\textwidth]{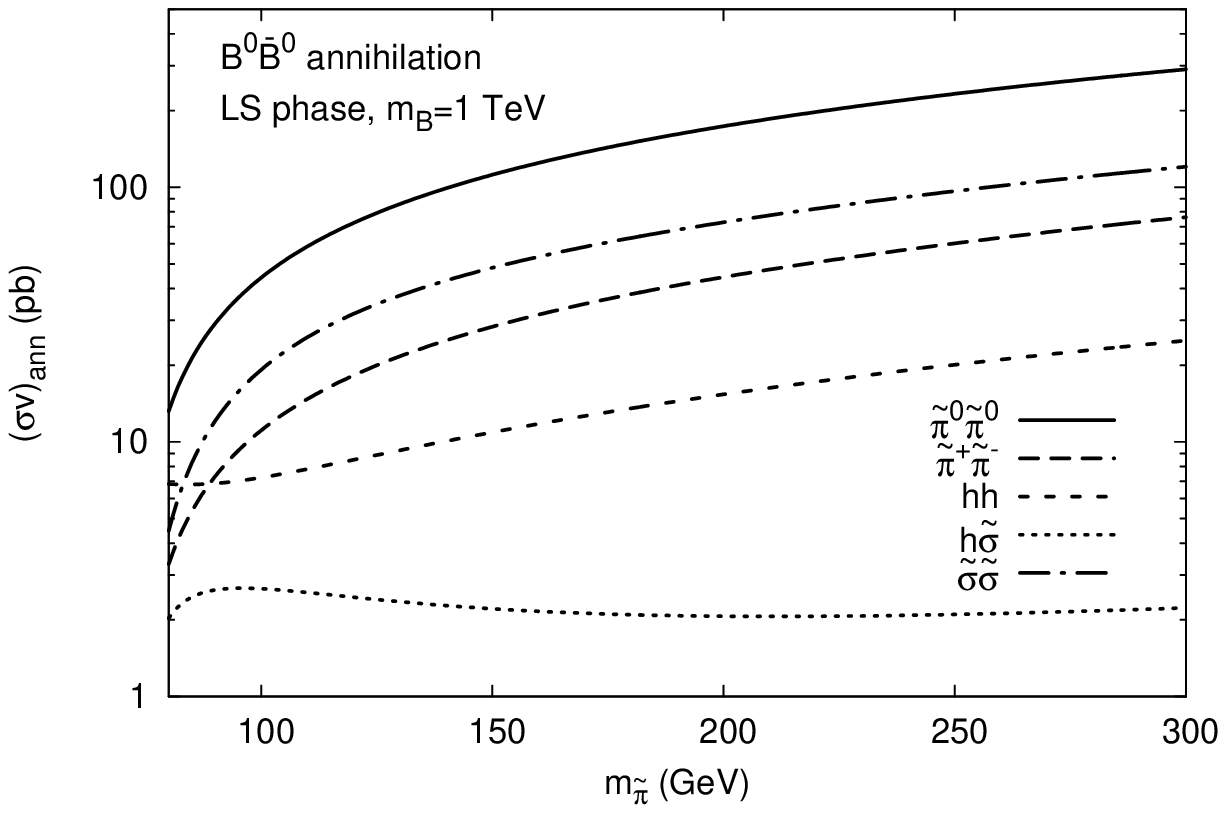}}
\end{minipage}
   \caption{The T-baryon LS annihilation cross sections in the (pseudo)scalar $B^0\bar B^0\to 
   \tilde{\pi}^0\tilde{\pi}^0,\tilde{\pi}^-\tilde{\pi}^-,hh,h\tilde{\sigma},\tilde{\sigma}\tilde{\sigma}$ 
   channels as functions of the T-pion mass $m_{\tilde \pi}$ for light $m_B=250$ GeV (left) and 
   heavy $m_B=1$ TeV (right) T-baryons. Here, $\Delta m_{\tilde{\sigma}}=20$ GeV is fixed.}
 \label{fig:TC-ann-Mpi}
\end{figure*}

It is worth to mention that the inequality (\ref{omega-TB}), 
$\Omega_{\rm TB}\lesssim \Omega_{\rm CDM}$, applied for the T-baryon
annihilation in the HS phase (\ref{HS-TC}) imposes a constraint on 
the chiral symmetry breaking scale
\begin{eqnarray} \label{u-constr}
u\gtrsim 4.3\, {\rm TeV} \gg v \,, \quad 
m_B\simeq 2u\,g_{\rm TC}\,,
\end{eqnarray}
which is valid under the naive scaling of the scalar self-couplings with 
the T-baryon mass (\ref{gB-coupl}). This constraint is consistent 
with the TC decoupling limit, and hence with the T-parameter
constraint and the SM-like Higgs boson as discussed above 
in Sect.~\ref{sec:VLC}. Together with Eq.~(\ref{u-constr}), 
the requirement for DM annihilation in the HS phase pushes up 
the T-baryon mass scale $m_B$ 
\begin{eqnarray}
T_f\simeq \frac{m_B}{20}\gtrsim T_{\rm EW}\,, \quad 
m_B\gtrsim 2\, {\rm TeV}\,,
\end{eqnarray}
leading to a simple lower bound on scalar self-interaction rates
\begin{eqnarray}
g_{{\rm B},i}=g_{\rm TC}\gtrsim g_{\rm TC}^{\rm min}\simeq \frac{1\,{\rm GeV}}{u}\,, 
\quad g_{\rm TC}^{\rm min}\lesssim 0.23\,.
\end{eqnarray}
Then, for the saturated inequality (\ref{omega-TB}) when 
all the DM is made of the scalar T-baryons only,
$\Omega_{\rm TB}\simeq \Omega_{\rm CDM}$, one obtains
\begin{eqnarray} \label{gTC-lower}
g_{{\rm B},i}=g_{\rm TC}\gtrsim 0.23\,.
\end{eqnarray}
This bound is consistent with the initial hypothesis about the weakly-interacting 
heavy T-baryons in the TC decoupling limit $\Lambda_{\rm TC}\gtrsim 
1\, {\rm TeV}\gg M_{\rm EW}$ formulated in Sect.~\ref{sec:TB-Lag}.
So, the heavy symmetric scalar T-baryon DM scenario with the relic abundance 
formation before the EW phase transition appears to be a feasible and appealing option.
Of course, for a more precise quantitative analysis one would need to know the 
exact dependence of non-perturbative T-baryon couplings on T-sigma vev $u$ 
(or, equivalently, on $\langle\bar{{\tilde Q}}{\tilde Q}\rangle$) going beyond 
the naive assumption (\ref{gB-coupl}) which is an important further subject 
for a lattice analysis (for a recent lattice study of effective Higgs--T-baryon 
interactions in $SU(4)$ gauge theory, see Ref.~\cite{Appelquist:2014dja})

The T-baryon annihilation in the {\it low-symmetry phase} (or LS phase) $B^0\bar B^0 \to WW$ 
which should be relevant for not very heavy particles $m_B\lesssim 1$ TeV, 
the weak-induced annihilation rate qualitatively change its energy dependence 
such that it does not disappear close to the threshold any longer. In opposite,
it becomes enhanced in the heavy T-baryon limit $m_B\gg m_W$, especially
in the case of a large $B^0-B^\pm$ mass splitting $\Delta m_B \gtrsim m_W$.
This effect can be seen in Fig.~\ref{fig:SM-ann}(left), where the $B^0\bar B^0\to WW$ cross section
is shown as a function of $\Delta m_B=[5\,\dots\, 100]$ GeV for three different mass
scale $m_B=0.2,1,3$ TeV. Indeed, in the small $\Delta m_B\to 0$ limit one 
observes gross cancellations between the contact and $t,u$-channel contributions 
($s$-channel terms are always suppressed) while at larger $\Delta m_B$ the 
cancellations are less precise leading to such a characteristic shape of the cross 
section. We can immediately conclude from this result that the case of symmetric T-baryon 
DM which undergoes its annihilation mainly in the LS phase of the cosmological 
plasma can not be realized for large T-baryon mass splittings $\Delta m_B\gtrsim 10$ 
GeV and relatively light T-baryons $m_B < 2$ TeV. This is thus the particular version of 
ADM whose relic abundance is fully characterized by the initial T-baryon asymmetry
providing a stable relic remnant of particles with the same T-baryon number
and no strict constraints on T-baryon interaction rates analogical to that in the HS phase 
(\ref{gTC-lower}) can be drawn in this case.

In Fig.~\ref{fig:SM-ann}(right) the largest T-baryon annihilation cross sections into SM 
particles $WW,ZZ,t\bar t$ are shown as functions of the T-baryon mass 
scale $m_B=[0.2\, \dots \, 3]$ TeV. Here, only $WW$ channel is sensitive 
to $\Delta m_B$ which in the latter figure has been fixed to $10$ GeV, 
for comparison.
\begin{figure*}[tbh]
 \centerline{\includegraphics[width=0.7\textwidth]{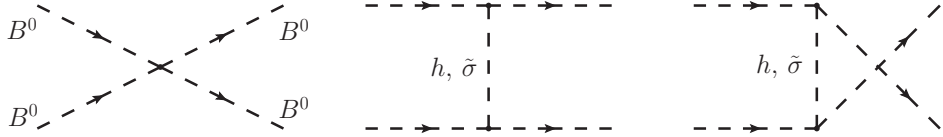}}
   \caption{
 \small Diagrams contributing to the elastic scattering of non-relativistic scalar T-baryons.}
 \label{fig:BB-BB-elastic}
\end{figure*}

The TC-induced annihilation channels into $h,\tilde\sigma,\tilde\pi$ 
final states approximately add up to
\begin{eqnarray} \label{LS-TC}
(\sigma v)^{\rm TC,LS}_{\rm ann} &\simeq& 
\frac{g_{\rm LS}^2}{16\pi m_B^2} \,,
\end{eqnarray}
altogether where in the case of large coupling $g_{\rm BS}\gg 1$ 
(or equivalently $m_B\gg u$) we have
\begin{eqnarray} \nonumber
g_{\rm LS}^2 &=& \Big(1-\frac{4m_h^2}{m_B^2}\Big)
\Big(\frac12 g_{\rm BH} c_\theta^2 + g_{\rm BS} s_\theta^2 \Big)^2 \\
\nonumber &+&  
\Big(1-\frac{4m_{\tilde \pi}^2}{m_B^2}\Big) \big( 2 g_{\rm BS}^2 + 
(g_{\rm BS} + g_{\rm BP})^2 \big) \\ 
\nonumber &+& 
\frac12\bar{\lambda}(m_h^2,m_{\tilde \sigma}^2;m_B^2)
(2g_{\rm BS}-g_{\rm BH})^2 s_\theta^2 c_\theta^2 \\ 
\nonumber &+&
\Big(1-\frac{4m_{\tilde \sigma}^2}{m_B^2}\Big)
\Big(g_{\rm BS} c_\theta^2 + \frac12 g_{\rm BH} s_\theta^2 \Big)^2 \,,
\end{eqnarray}
corresponding to the contact diagrams in Fig.~\ref{fig:TB-annihilation} only. 
The exact result applicable also for the case of small effective couplings, e.g. 
$g_{\rm BS}\lesssim 1$, and hence for relatively light T-baryons $m_B\lesssim u$, 
is more complicated since it includes additional $s,t,u$-channel diagrams; it can be found 
in Appendix \ref{app}.
Note, in the small $h\tilde \sigma$-mixing limit $s_\theta\ll 1$,
or $u\gg v$, and heavy T-baryon limit $m_B\gg u$, or $g_{\rm TC}\gg 1$,
\begin{eqnarray} \nonumber
g_{\rm LS}\simeq g_{\rm HS}\equiv g_{\rm eff}\,.
\end{eqnarray}
as expected. 

The $B^0\bar B^0$ annihilation cross sections in the TC-induced (pseudo)scalar 
$\tilde{\pi}^0\tilde{\pi}^0,\tilde{\pi}^-\tilde{\pi}^-,hh,h\tilde{\sigma},\tilde{\sigma}\tilde{\sigma}$ 
channels are presented in Fig.~\ref{fig:TC-ann-mB_DMS} as functions of the T-baryon mass scale $m_B$ 
(left panel) and $\Delta m_{\tilde{\sigma}}\equiv m_{\tilde \sigma}-\sqrt{3}m_{\tilde{\pi}}$ (right panel).
The $m_B$ dependence flattens out at large $m_B$ except for the $h\tilde{\sigma}$ channel 
since $g_{{\rm B},i}=g_{\rm TC}\sim m_B$ (c.f. Eq.~\ref{gB-coupl}). In the TC decoupling limit $u\gg v$,
corresponding to $s_\theta\ll 1$ and hence to $\Delta m_{\tilde{\sigma}}\to 0$, all the 
(pseudo)scalar annihilation channels vanish which is an important peculiarity of the considering scenario since 
$g_{{\rm B},i}=g_{\rm TC}\sim 1/u \to 0$ while there is no exact symmetry of the cross 
sections w.r.t. $\Delta m_{\tilde{\sigma}}\leftrightarrow -\Delta m_{\tilde{\sigma}}$. In Fig.~\ref{fig:TC-ann-Mpi} 
the same cross sections are plotted as functions of the T-pion mass $m_{\tilde \pi}$ for
two different T-baryon mass values $m_B=250$ GeV (left panel) and 1 TeV (right panel). 
Consequently, one finds a disappearance of the total TC-induced cross section 
in the heavy T-pion limit $m_{\tilde \pi}\gtrsim m_B$ which is otherwise peaked 
in intermediate regions.
\begin{figure}[h!]
 \centerline{\includegraphics[width=0.5\textwidth]{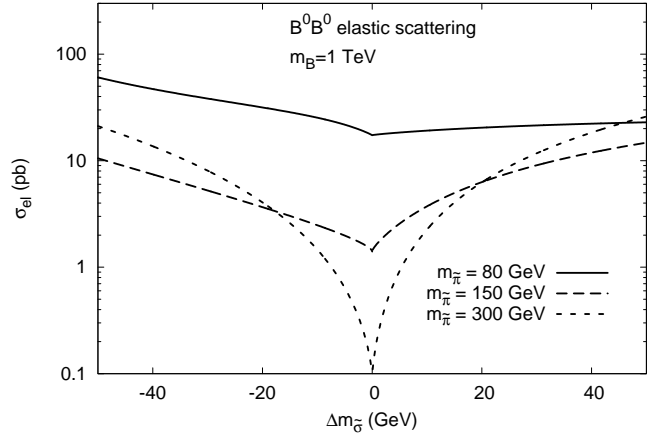}}
   \caption{
 \small The T-baryon elastic scattering cross section $\sigma_{el}$ as a function 
 of $\Delta m_{\tilde{\sigma}}$ for three different values of the 
 T-pion mass $m_{\tilde \pi}=80,150$ and 300 GeV. 
 Here, $m_B=1$ TeV is fixed.}
 \label{fig:BB-BB-elastic-CS}
\end{figure}

It is worth to summarize that the total T-baryon annihilation rate is 
generally lower in the HS phase than that in the LS one not only by 
means of a higher T-baryon mass scale $m_B$, but also due to vanishing weak-induced 
contribution (\ref{HS-W-v}). In opposite, the weak-induced channels,
especially into $WW$, appear to dominate the T-baryon annihilation 
cross section in the LS phase for a not too small T-baryon mass splitting.
The bottom line of this study is that it is impossible to accommodate 
the symmetric T-baryon DM in the LS annihilation scenario unless the
TC decoupling limit $\Delta m_{\tilde{\sigma}}\to 0$ and, simultaneously, the 
heavy T-baryon and low mass splitting limits are realized effectively approaching 
the HS annihilation scenario developed in Eqs.~(\ref{HS-W})--(\ref{gTC-lower}).
Therefore, we encounter two rather different HS/LS annihilation scenarios which 
thus lead to symmetric/asymmetric T-baryon DM, respectively. The DM of an intermediate 
or mixed type can also be accommodated in the HS phase depending on the amount of
initial T-baryon asymmetry. 

\subsection{Elastic T-baryon scattering}

One of the important inputs for the cosmogonic DM evolution and
Structure Formation in late Universe is the elastic WIMP-WIMP
scattering cross section. In the case of non-relativistic T-baryon
DM, the elastic $B^0B^0\to B^0B^0$ scattering is described by five
diagrams -- one contact diagram, two $t$-channel and corresponding 
cross ($u$-channel) diagrams via $h$ and $\tilde \sigma$ exchanges 
shown in Fig.~\ref{fig:BB-BB-elastic}. The non-relativistic T-baryon elastic 
scattering cross section is given by
\begin{eqnarray} \label{BB-el}
 \sigma_{el}\simeq\frac{G_{\rm B}^2}{32\pi m_B^2}\,, \;\,
 G_{\rm B}\equiv g_{\rm BB}+8u^2g_{\rm BS}^2
 \Big(\frac{c_\theta^2}{m_{\tilde \sigma}^2}+\frac{s_\theta^2}{m_h^2}\Big)\,.
\end{eqnarray}
As a reference estimate, for $G_B\sim 1-10$ and 1 TeV T-baryon,
one obtains $\sigma \sim 15\times (1-100)$ pb, respectively.
In Fig.~\ref{fig:BB-BB-elastic-CS} the elastic cross section is illustrated
as a function of $\Delta m_{\tilde{\sigma}}$ under the
setting (\ref{gB-coupl}). So, the elastic T-baryon scattering appears 
to be weaker than usual elastic nucleon scattering but significantly 
stronger than DM particles' scattering in ordinary WIMP-based scenarios.
Therefore, in fact we deal with a particular case of self-interacting DM which
may be useful for DM astrophysics (see e.g. Ref.~\cite{Spergel:1999mh}).

\section{Dark Matter detection prospects}

\subsection{Indirect detection}

At later stages of the Universe evolution, namely, after termination of T-baryon annihilation 
epoch, the T-baryon interactions are described by the effective low-energy Lagrangians (\ref{LsM}) 
and (\ref{Delta-LB}) where the chiral and EW symmetries breaking should be performed 
according to Eqs.~(\ref{shifts}) and (\ref{u-v-min}). In the case of {\it symmetric} 
T-baryon DM $m_B\gtrsim 2$ TeV, its late time LS annihilation is described by the cross sections 
shown in Figs.~\ref{fig:SM-ann}, \ref{fig:TC-ann-mB_DMS} and \ref{fig:TC-ann-Mpi}. Here, one could 
distinguish two distinct cases: small $\Delta m_B < 10$ GeV and large $\Delta m_B > 10$ GeV 
T-baryon mass splitting. 
\begin{figure}[!h]
 \centerline{\includegraphics[width=0.22\textwidth]{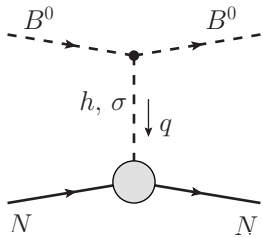}}
   \caption{
 \small The elastic scattering of neutral scalar T-baryon $B^0$ off a
 nucleon target $N$ at tree level via $h$ and $\sigma$ exchanges 
 in the $t$-channel relevant for direct DM detection close 
 to the threshold $\sqrt{s_{\rm th}}\equiv m_B+m_N$.}
 \label{fig:TB-elastic-tree}
\end{figure}

In the first case, the $B^0\bar B^0\to WW$ annihilation channel is suppressed, and $WW,\,ZZ$ and 
$t\bar t$ channels altogether contribute at most $\sim 10$ pb or less to the total cross section,
which is therefore dominated by T-strong channels for not too small $\Delta m_{\tilde\sigma}\gtrsim 5$ GeV, 
specifically, by
\[
B^0\bar B^0\to \tilde\pi^0 \tilde\pi^0\,,\tilde\pi^- \tilde\pi^-\,,\tilde\sigma\tilde\sigma \,,
\]
whose total contribution exceeds e.g. $100$ pb for $m_B=2$ TeV and $\Delta m_{\tilde\sigma}=20$ GeV.
Then, the dominant decay modes of T-sigma and T-pions with rather large (over $90$ \%) branching
ratios in the considering 2-flavor $SU(2)_{\rm TC}$ VLC scenario are
\[
\tilde\sigma\to \tilde\pi^{0,\pm}\tilde\pi^{0,\mp}\,,\quad \tilde\pi^0 \to Z\gamma\gamma\,, 
\quad \tilde\pi^\pm \to W^\pm \gamma\gamma\,,
\]
respectively. We conclude that a key distinct indirect signature of the symmetric scalar T-baryon 
DM at a TeV mass scale will be its annihilation into predominantly light T-pions decaying into a vector
boson and into two energetic photons which is a straightforward subject for a promising multi-GeV
$\gamma$-lines search at FERMI \cite{Ackermann:2013uma}. Finally, in the very strong TC decoupling limit 
when $\Delta m_{\tilde\sigma}\ll 5$ GeV, the only $B^0\bar B^0\to W^\pm W^\mp$ annihilation 
channel survives while the $ZZ$ and $f\bar f$ ones vanish since the T-baryon--Higgs boson 
coupling is expected to decrease with the TC scale, e.g. $g_{\rm BH}\sim 1/u$ under 
the setting (\ref{gB-coupl}). The similar indirect observational consequences can be drawn
for the T-baryon DM of a mixed type, with a lower density of particles and antiparticles 
capable of mutual annihilation, which may diminish corresponding detection rates. This situation
means that even in the highly theoretically constrained VLC scenario, there is phenomenologically 
enough freedom to accommodate the heavy scalar symmetric T-baryon DM scenario predicting
rather interesting and potentially detectable hard multi-$\gamma$ signatures. 

As was advocated in the previous section, if the T-baryon mass splitting is 
large $\Delta m_B > 10$ GeV and/or the mass scale is lower $m_B < 2$ TeV, 
one deals with the purely asymmetric T-baryon DM (other possibilities would be 
highly fine-tuned and are thus less likely). This is a pessimistic scenario for indirect 
DM detection measurements looking for T-baryon annihilation products -- in the 
absence of a significant amount of anti-T-baryons and exact T-baryon conservation 
in present Universe, one does not expect to see any DM annihilation signatures. Only,
direct ADM detection is relevant, so does in the symmetric DM case, considered above.

\subsection{Direct detection}

Consider now the scalar T-baryon implications for the direct DM detection
experiments looking for nuclear recoils due to elastic
non-relativistic WIMP-nucleon scattering. At {\it tree level}, the elastic
scalar T-baryon--nucleon scattering is mediated by the scalar Higgs
boson and T-sigma exchanges in the $t$-channel only as shown in
Fig.~\ref{fig:TB-elastic-tree}. 
\begin{figure*}[tbh]
 \centerline{\includegraphics[width=0.75\textwidth]{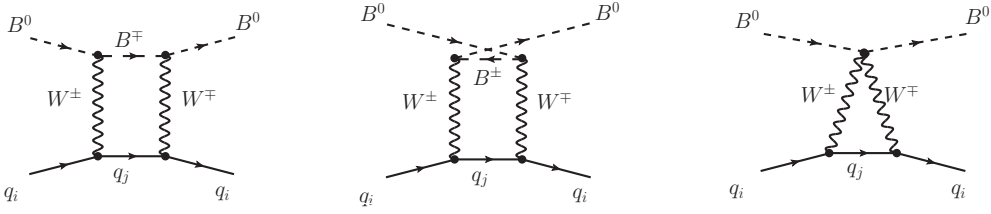}}
   \caption{
 \small The elastic scattering of neutral scalar T-baryon $B^0$ off a
 light quark target $q_i$ at one loop level. Only the dominant
 topologies are shown.}
 \label{fig:TB-elastic-loop}
\end{figure*}

Consider the consistent TC decoupling limit, 
$s_\theta\ll 1$. In this case, the elastic scattering of non-relativistic 
T-baryons off nucleons in an underground DM detector is dominated 
essentially by the Higgs boson exchange with a very small
$q^2\sim 1\,{\rm KeV}^2$. For such small $q^2$ values (slightly above
the threshold $\sqrt{s_{\rm th}}\equiv m_B+m_N$)
one deals the effective Higgs-nucleon form factor previously discussed 
in Ref.~\cite{Shifman:1978zn}. Then, the effective spin-independent 
T-baryon--nucleon elastic cross section reads
\begin{eqnarray} \nonumber
\sigma^{\rm nucl}_{\rm SI} &\simeq&
\frac{\kappa^2}{\pi}\frac{m_N^4}{m_B^2m_h^4}\,F_N^2 \\ 
&\simeq& 5.45\,\kappa^2\Big(\frac{1\,\rm{GeV}}{m_B}\Big)^2\cdot
10^{-38}\;\rm{cm}^2\,, \label{SI}
\end{eqnarray}
where $m_N$ is the nucleon mass, and $F_N$ parameterizes the
effective tree-level matrix element for the Higgs interaction with
the nucleon target incorporating the gluon anomaly \cite{Shifman:1978zn,Giedt:2009mr}
\begin{eqnarray}
 && \frac{1}{v}\,\langle N|\sum_q m_q\bar{q}q|N \rangle\equiv F_N\,
 \frac{m_N}{v}\,\langle NN \rangle\,.
\end{eqnarray}
In Eq.~(\ref{SI}), we used the SM lattice result $F_N\simeq 0.375$ 
following the corresponding analysis of Ref.~\cite{Hill:2011be} where 
the uncertainty is dominated by the strange quark contribution, and $\kappa$ 
is the effective T-baryon--Higgs coupling which in the generic VLC scenario reads
\begin{eqnarray} \nonumber
{\cal L}_{BBh} = \kappa v\, B\bar B h\,, \quad \kappa \simeq g_{\rm
BH}+\frac{2m_h^2}{m_{\tilde \sigma}^2-m_h^2}\,g_{\rm BS}\,.
\end{eqnarray}

At {\it one-loop level}, one encounters an appearance of gauge boson
mediated contributions to the elastic T-baryon scattering off a
quark in the nucleon target. Thus, it is meaningful to consider first 
the elementary $B^0 q\to B^0 q$ subprocess for such 
a short-range loop-induced reaction.
\begin{figure}[!h]
 \centerline{\includegraphics[width=0.47\textwidth]{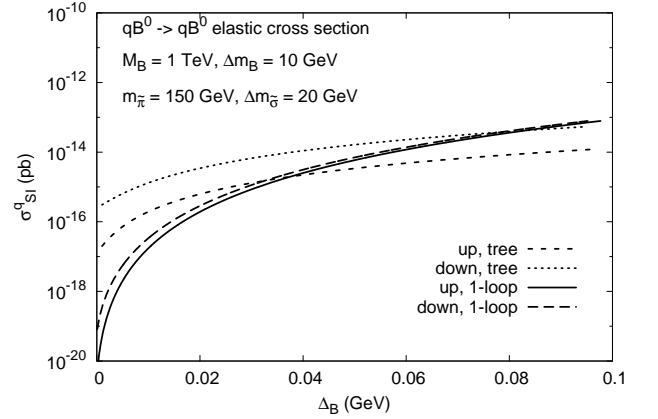}}
   \caption{
 \small The elastic T-baryon--quark scattering cross section
 of neutral scalar T-baryon $B^0$ off a light ($u,d$) quark
 target at one loop level as a function of $\Delta_B$. The
 corresponding tree-level cross section is shown for comparison.}
 \label{fig:TB-elastic-tree-vs-loop}
\end{figure}

We have performed a dedicated analysis of the full one-loop order
correction to the $B^0q\to B^0q$ ($q=u,d$) elastic cross section close to
the corresponding threshold $\sqrt{s_{\rm th}}$. The dominant diagrams are
shown in Fig.~\ref{fig:TB-elastic-loop}, while other possible topologies are found
to be strongly suppressed. For this purpose, the effective Lagrangian of T-baryon
interactions (\ref{Delta-LB}) together with the VLC model (\ref{LsM}) has been 
implemented into the FeynRules framework \cite{Alloul:2013bka} whose output was then 
used by FeynArts \cite{Hahn:2000kx} to calculate the respective one-loop amplitudes. 
The latter were processed by the FormCalc package \cite{Hahn:1998yk} into a Fortran code, 
together with necessary VLC parameter relations and mass formulae. The one-loop
master integrals were evaluated by the LoopTools package \cite{Hahn:1998yk}, and
the final cross section has been evaluated for relevant sets of physical parameters.
Namely, the one-loop correction to the elastic $B^0q\to B^0q$ scattering
cross section has been found as a function of
\begin{eqnarray}
 \Delta_B\equiv \sqrt{s}-\sqrt{s_{\rm th}} =
 \frac{2m_N}{m_B}\,E_B^{\rm lab}\,,
\end{eqnarray}
where $E_B^{\rm lab}=m_Bv^2/2$ ($v\sim 0.001$) is the T-baryon
kinetic energy in the laboratory frame. The result shown in
Fig.~\ref{fig:TB-elastic-tree-vs-loop} clearly demonstrates
a strong suppression of the one-loop correction compared 
to tree-level $B^0q\to B^0q$ scattering cross section in the non-relativistic
scalar T-baryon limit, very close to the threshold. This is the case of thermal relic 
T-baryons in a Sun neighborhood. Namely, individual $u$ and $d$ quark
scatterings at one loop level are found to be over three orders of magnitude smaller
then the corresponding tree-level results. Therefore, the cross
section given by $t$-channel Higgs boson exchange (\ref{SI}) is considered
to be accurate enough and sufficient for a comparison to the DM direct experimental constraints.
Note, for energetic T-baryons away from the threshold the one-loop induced cross
section becomes large and dominate the elastic T-baryon scattering off a quark.

\begin{figure}[!h]
 \centerline{\includegraphics[width=0.47\textwidth]{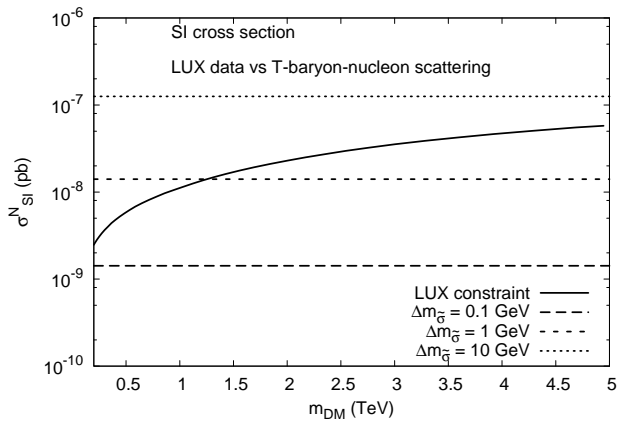}}
   \caption{
 \small The spin-independent T-baryon--nucleon cross section vs LUX constraint \cite{Akerib:2013tjd}.}
 \label{fig:TB-elastic-vs-LUX}
\end{figure}

In Ref.~\cite{Appelquist:2014dja}, the effective scalar T-baryon--Higgs coupling has
been constrained by lattice simulations in the strongly-coupled
$SU(4)_{\rm TC}$ model. While a similar analysis in the VLC model
with confined $SU(2)_{\rm TC}$ symmetry yet has to be performed, in
this first study we treat $\kappa$ as a free parameter which can be
constrained by the direct DM detection data, together with scalar
T-baryon mass scale $m_B$. At the moment, the LUX experiment
\cite{Akerib:2013tjd} provides the most stringent limit on $\sigma^{\rm
nucl}_{\rm SI}$ (per nucleon in the case of a xenon target) which is
roughly
\begin{eqnarray}
-\log_{10}\Big(\frac{\sigma^{\rm nucl}_{\rm SI}}{{\rm
cm}^2}\Big)\simeq 44.2 - 43.7\,, \;\, m_{\rm B}\simeq 0.5 - 2\, {\rm
TeV}\,,
\end{eqnarray}
providing the following bound on effective coupling
\begin{eqnarray}
\kappa \lesssim 0.17-1.2\,, \label{bound}
\end{eqnarray}
respectively, and somewhat weaker constraints for larger $m_B$.
This is in agreement with the lower bound on T-strong couplings from the heavy 
symmetric T-baryon constraint (\ref{gTC-lower}) and with our expectations 
for relatively weak T-baryon--T-meson interactions in the small $h\sigma$-mixing limit
noticed above. Indeed, in Fig.~\ref{fig:TB-elastic-vs-LUX} the recent 
LUX bound has been plotted vs T-baryon mass scale $m_B$ together 
with the effective scalar T-baryon--nucleon SI cross section given 
by Eq.~(\ref{SI}). The latter have been obtained for three distinct 
$\Delta m_{\tilde \sigma}$ values by using the naive scaling 
condition for scalar TC couplings (\ref{gB-coupl}). The cross section
is less sensitive to the T-pion mass so it has been fixed 
to $m_{\tilde \pi}=150$ GeV. 

We note that the direct DM detection constraint from LUX in practice excludes large
$\Delta m_{\tilde \sigma}\gtrsim 10$ GeV. It leaves the space
only for a narrow a few GeV region in the vicinity of the TC 
decoupling limit $\Delta m^{dec}_{\tilde \sigma}\equiv 0$. Therefore, the
direct DM constraint is complimentary to the corresponding
constraints on $h\tilde\sigma$ mixing angle from the Higgs decays 
and from the T-parameter. It is worth to stress that the DM direct 
bound in Fig.~\ref{fig:TB-elastic-vs-LUX} has turned out to be 
the most stringent constraint on new $SU(2)$ confined dynamics 
among other ones. Indeed, it pushes the acceptable $SU(2)_{\rm TC}$ scale 
$\Lambda_{\rm TC}$ even further away from the EW one (in a few TeV region) 
then the EW precision constraints. The bound (\ref{bound}) should be tested against
lattice simulations in the considered strongly-coupled $SU(2)_{\rm
TC}$ theory.

\subsection{Collider signatures}

The composite DM studies, and the new $SU(2)_{\rm TC}$ strong dynamics searches, 
in general, should be accompanied by respective scalar T-baryon searches at the LHC.
In order to digest this possibility, let us briefly discuss the corresponding signatures 
of scalar T-baryon production in $pp$ collisions at $14$ TeV. 
\begin{figure}[!h]
 \centerline{\includegraphics[width=0.47\textwidth]{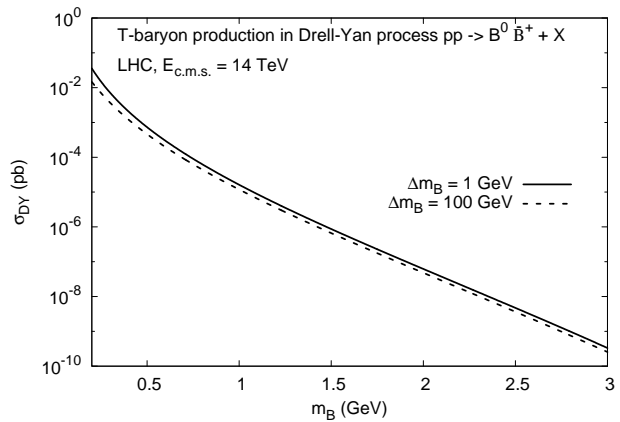}}
   \caption{
 \small The T-baryon $B^0\bar{B}^+$ pair production cross section 
 via Drell-Yan channel in $pp$ collisions at the LHC, $E_{c.m.s.}=14$ TeV, for
 two distinct T-baryon mass splitting values, $\Delta m_B=1$ and 100 GeV.}
 \label{fig:TB-LHC}
\end{figure}

One of the characteristic signals can be expected from the Drell-Yan
charged-current reaction 
\[
q_i\bar q_j \to (W^\pm)^* \to B^0 \bar B^\pm\,, \quad 
\bar B^\pm \to \bar B^0(W^\pm\to f_i \bar f_j)\,, 
\]
with deeply virtual $W$ boson in the $s$ channel. Here, heavy $B^0$ and $\bar B^0$
leave the detector unnoticed giving rise to a very large (a few TeV) missing mass in the
missing $E_T^{miss}$ spectrum, while the tagging on charged debris from the 
final $W$ decay may be very problematic for a relatively small mass splitting 
$\Delta m_B\lesssim 50-70$ GeV. Also, this is due to a very dramatic dependence on the
$B^\pm$ width on $\Delta m_B$ shown in Fig.~\ref{fig:TB-decay}, such that at small 
$\Delta m_B$ the charged $B^\pm$ ``lives'' longer before it decays into a $W$ and invisible 
$B^0$ (a possibility for a displaced $W$ emission accompanied by heavy invisibles). Note, 
the corresponding $B^0 \bar B^\pm$ production cross section shown 
in Fig.~\ref{fig:TB-LHC} depends strongly on mass scale $m_B$ only and does not 
depend on any other TC parameters and couplings, and only weakly depends on 
the T-baryon mass splitting $\Delta m_B$. Further more detailed studies of the T-baryon detection
capabilities including realistic detector constraints should be performed elsewhere.

\section{Summary and conclusions}

In this work, we have performed a detailed theoretical and phenomenological 
study of scalar T-baryon sector, in particular, with respect to its possible important
role for DM astrophysics. The corresponding scenario of new gauge dynamics is based
upon the consistent vector-like TC model with $SU(2)_{\rm TC}$ group which
allows to obtain Dirac T-quarks in confinement from original chiral fermion multiplets. Starting from 
the phenomenological arguments provided by (i) EW precision tests and (ii) SM Higgs-like
couplings, the T-baryon states turn out to be split and pushed up towards higher scales, away
from the dynamical EW symmetry breaking scale (the TC decoupling limit, $\Lambda_{\rm TC}\gg M_{\rm EW}\sim 100$ GeV). 
In this regime, the scalar neutral T-baryon (T-diquark) $B^0=UD$ state (the lightest among other T-diquarks) 
possessing a conserved quantum number can serve as an appealing DM candidate at a TeV mass scale, 
$m_B\gtrsim 0.5-1$ TeV. In the consistent TC decoupling limit, the scalar self-couplings of heavy T-baryons 
with light T-pions/T-sigma states are expected to decrease. 

For $m_B>2$ TeV one expects that the relic T-baryon abundance has been formed mainly before the 
EW phase transition. We have shown that in this case the total T-baryon annihilation cross section 
is rather weak and enables symmetric DM formation, whereas lighter T-baryons annihilating mostly 
in the low-symmetry phase of the cosmological plasma could only remain if there was a significant 
T-baryon asymmetry generation. A specific astrophysical signature of symmetric T-baryon DM 
is its annihilation into hard multi-$\gamma$ final states via intermediate T-pion/T-sigma states, 
relevant for indirect DM detection measurements. 

We have shown that the elastic T-baryon--quark scattering is induced at tree level by the Higgs/T-sigma 
bosons exchanges in the $t$-channel only, since the vector tree-level $Z$-boson is absent. 
We have calculated complete one-loop correction to the elastic
scalar T-baryon scattering and found that it is strongly suppressed compared to the tree-level result. 
Most importantly, the direct DM detection constraints, e.g. those from LUX, on spin-independent elastic 
T-baryon--nucleon scattering cross section impose further ever stringent constraint on the T-baryon--Higgs coupling 
and hence on the chiral symmetry breaking scale. Namely, it imposes a much stronger suppression to the $h\tilde\sigma$ 
mixing angle (and hence the stronger hierarchy between the chiral and EW breaking scales) than the corresponding 
EW precision and SM Higgs bounds.

Finally, the search for heavy scalar T-baryons in the Drell-Yan production process with a trigger on a few TeV missing
mass and, possibly, on accompanying $W$ emission from a displaced vertex, is advised. A further more dedicated
analysis of T-baryon implications at the LHC, in particular, in vector-boson fusion channels would be desirable.

Note, the question about compositeness of the light Higgs boson is not critical for cosmological evolution
of the T-baryon DM analysis in the TC decoupling limit, $\Lambda_{\rm TC}\gg 100$ GeV, when the light 
technipion limit $m_{\tilde \pi}\ll \Lambda_{\rm TC}$ is concerned as long as scalar T-baryon--Higgs 
boson coupling is within the perturbative limit. Indeed, as we have advocated above in detail the couplings 
of heavy T-baryons with light composites (e.g. with pseudo-Goldstone states) are suppressed in the decoupling limit. 
As we have demonstrated above, the existing direct DM detection bounds set a limit on the T-baryon-Higgs 
boson coupling which should be understood as an important constraint on the existing variety of composite Higgs 
models. An account for the Higgs compositeness will not change this situation and, specifically, does not affect 
our basic conclusions neither for asymmetric, nor for symmetric DM scenarios. This is a motivation for the study 
of the cosmological T-baryon evolution independently on aspects of possible Higgs compositeness which will 
be thoroughly discussed in our forthcoming works. It is worth to stress here that the LUX experiment bound 
\cite{Akerib:2013tjd} in the TC decoupling limit is consistent with cosmological evolution and, in particular, 
with the freeze out of heavy T-baryons in the high-symmetry phase of the cosmological plasma setting 
an important stage for further development of composite DM scenarios in the TC decoupling limit.

An incorporation of the Sommerfeld enhancement effect into the cosmological evolution of scalar T-baryons
could be a nice development of this work. However, in this first study we prefer not to incorporate this
effect and hence to introduce an extra freedom into our analysis naively assuming a sharp cutoff in self-annihilation 
rates after the T-baryon freeze-out temperature. In fact, it is believed that the freeze out happens at relative 
velocities of about $v\sim 0.1-0.3$ where the Sommerfeld enhancement factor is of the order of unity for 
rather weak T-baryon-Higgs couplings suggested by both the TC decoupling limit and direct DM detection 
constraints. A proper analysis of the Sommerfeld enhancement effect could be a good point for further studies. 

\vspace{0.5cm}
 {\bf Acknowledgments}

Stimulating discussions and helpful correspondence with Johan Bijnens, 
Johan Rathsman and Torbj\"orn Sj\"ostrand are acknowledged.
V. B. and V. K. were partially supported by Southern Federal University 
grant No. 213.01-.2014/-013VG. R. P. is grateful to the ``What is Dark Matter?'' 
Program at Nordita (Stockholm) for support and hospitality during 
completion of this work. R. P. was supported in part by the Swedish 
Research Council, contract number 621-2013-428.

\appendix

\section{T-baryon annihilation into T-mesons}
\label{app}

In Eq.~(\ref{LS-TC}) only contact diagrams dominating in the limit $m_B\gg u$ 
have been included. Here, we list the complete results for partial annihilation 
cross sections including all diagrams in Fig.~\ref{fig:TB-annihilation} and
hence applicable for any hierarchy between $m_B$ and 
$u,\, m_h,\, m_{\tilde\sigma}$. These read
\begin{eqnarray}
(\sigma v)^{\rm XY,LS}_{\rm ann} \simeq 
\frac{g_{\rm XY}^2}{16\pi m_B^2} \,\bar{\lambda}(m_X^2,m_Y^2,m_B^2) \,,
\end{eqnarray}
where $\bar{\lambda}(a,b,c)$ is defined in Ref.~(\ref{lambda}), and effective
couplings are
\begin{eqnarray*}
g_{hh} &=& \frac12 g_{\rm BH} c_\theta^2 + g_{\rm BS} s_\theta^2 \\
&+& \frac{2}{s-2m_h^2}\big(2u\, g_{\rm BS} s_\theta + v\, g_{\rm BH} 
c_\theta \big)^2 \\ 
&-& \frac{g_{{\tilde\sigma}hh}}{s-m_{\tilde\sigma}^2}
\big(2u\, g_{\rm BS} c_\theta - v\, g_{\rm BH} s_\theta \big) \\ 
&-& \frac{g_{hhh}}{s-m_h^2}\big(2u\, g_{\rm BS} s_\theta + 
v\, g_{\rm BH} c_\theta \big) \,, \\
g_{{\tilde \sigma}{\tilde \sigma}} &=& \frac12 g_{\rm BH} s_\theta^2 
+ g_{\rm BS} c_\theta^2 \\
&+& \frac{2}{s-2m_{\tilde \sigma}^2}\big(2u\, g_{\rm BS} c_\theta 
- v\, g_{\rm BH} s_\theta \big)^2 \\ 
&-& \frac{g_{h{\tilde\sigma}{\tilde\sigma}}}{s-m_h^2}
\big(2u\, g_{\rm BS} s_\theta + v\, g_{\rm BH} c_\theta \big) \\
&-& \frac{g_{{\tilde\sigma}{\tilde\sigma}{\tilde\sigma}}}
{s-m_{\tilde\sigma}^2}\big(2u\, g_{\rm BS} c_\theta - 
v\, g_{\rm BH} s_\theta \big)\,, \\
\sqrt{2}\,g_{{\tilde \sigma}h} &=& (2g_{\rm BS}-g_{\rm BH}) 
s_\theta c_\theta \\
&+& \frac{4(s-m_h^2-m_{\tilde \sigma}^2)}{(s-2m_h^2)
(s-2m_{\tilde \sigma}^2)}\big(2u\, g_{\rm BS} s_\theta + 
v\, g_{\rm BH} c_\theta \big) \\ &\times & 
\big(2u\, g_{\rm BS} c_\theta - v\, g_{\rm BH} s_\theta \big) \\
&-& \frac{g_{h{\tilde\sigma}{\tilde\sigma}}}{s-m_{\tilde \sigma}^2}
\big(2u\, g_{\rm BS} c_\theta - v\, g_{\rm BH} s_\theta \big) \\
&-& \frac{g_{{\tilde\sigma}hh}}{s-m_h^2}
\big(2u\, g_{\rm BS} s_\theta + v\, g_{\rm BH} c_\theta \big) \\
g_{{\tilde \pi}^0{\tilde \pi}^0} &=& g_{\rm BS}+g_{\rm BP} - 
\frac{g_{{\tilde\sigma}{\tilde\pi}^0{\tilde\pi}^0}}{s-m_{\tilde \sigma}^2}
\big(2u\, g_{\rm BS} c_\theta - v\, g_{\rm BH} s_\theta \big) \\
&-& \frac{g_{h{\tilde\pi}^0{\tilde\pi}^0}}{s-m_h^2}
\big(2u\, g_{\rm BS} s_\theta + v\, g_{\rm BH} c_\theta \big) \,, \\
\sqrt{2}\,g_{{\tilde\pi}^+{\tilde\pi}^-} &=& 2g_{\rm BS} - 
\frac{g_{{\tilde\sigma}{\tilde\pi}^+{\tilde\pi}^-}}{s-m_{\tilde \sigma}^2}
\big(2u\, g_{\rm BS} c_\theta - v\, g_{\rm BH} s_\theta \big) \\
&-& \frac{g_{h{\tilde\pi}^+{\tilde\pi}^-}}{s-m_h^2}
\big(2u\, g_{\rm BS} s_\theta + v\, g_{\rm BH} c_\theta \big) \,,
\end{eqnarray*}
where $s\simeq 4m_B^2(1+v^2/4)$ close to the threshold $v\ll 1$.



\begin{thebibliography}{99}

\bibitem{Aad:2012tfa} 
  G.~Aad {\it et al.}  [ATLAS Collaboration],
  Phys.\ Lett.\ B {\bf 716}, 1 (2012)
  [arXiv:1207.7214 [hep-ex]].

\bibitem{Chatrchyan:2012ufa} 
  S.~Chatrchyan {\it et al.}  [CMS Collaboration],
  Phys.\ Lett.\ B {\bf 716}, 30 (2012)
  [arXiv:1207.7235 [hep-ex]].

\bibitem{SM-tests}
  ATLAS Collaboration, ATLAS-CONF-2014-009 and ATLAS-CONF-2014-010, CERN Geneva, March 2014 ; \\
  S.~Chatrchyan {\it et al.}  [CMS Collaboration],
  arXiv:1401.6527 [hep-ex].
  
\bibitem{Chatrchyan:2013lba} 
  S.~Chatrchyan {\it et al.}  [CMS Collaboration],
  JHEP {\bf 1306}, 081 (2013)
  [arXiv:1303.4571 [hep-ex]].

\bibitem{TC-1}
S.~Weinberg, Phys. Rev. {\bf D13}, 974 (1976).

\bibitem{TC-2}
L.~Susskind, Phys. Rev. {\bf D20}, 2619 (1979).

\bibitem{Extended-TC-1}
  S.~Dimopoulos and L.~Susskind,
  Nucl.\ Phys.\ {\bf B155}, 237 (1979).

\bibitem{Extended-TC-2}
  E.~Eichten and K.~D.~Lane,
  Phys.\ Lett.\ {\bf B90}, 125 (1980).

\bibitem{Hill:2002ap}
  C.~T.~Hill and E.~H.~Simmons,
  Phys.\ Rept.\  {\bf 381}, 235 (2003)
  [Erratum-ibid.\  {\bf 390}, 553 (2004)]
  [hep-ph/0203079].

\bibitem{Sannino}
F.~Sannino,
  Acta Phys.\ Polon.\ B {\bf 40}, 3533 (2009).

\bibitem{Vecchi:2013bja} 
  L.~Vecchi,
  arXiv:1304.4579 [hep-ph].

\bibitem{Barducci:2013wjc} 
  D.~Barducci, A.~Belyaev, M.~S.~Brown, S.~De Curtis, S.~Moretti and G.~M.~Pruna,
  JHEP {\bf 1309}, 047 (2013)
  [arXiv:1302.2371 [hep-ph]].

\bibitem{DeSimone:2012fs} 
  A.~De Simone, O.~Matsedonskyi, R.~Rattazzi and A.~Wulzer,
  JHEP {\bf 1304}, 004 (2013)
  [arXiv:1211.5663 [hep-ph]].

\bibitem{Peskin-1}
  M.~E.~Peskin and T.~Takeuchi,
  Phys.\ Rev.\ Lett.\  {\bf 65}, 964 (1990).

\bibitem{Peskin-2}
  M.~E.~Peskin, T.~Takeuchi,
  Phys.\ Rev.\ D {\bf 46}, 381 (1992).

\bibitem{Kilic:2009mi} 
  C.~Kilic, T.~Okui and R.~Sundrum,
  JHEP {\bf 1002}, 018 (2010)
  [arXiv:0906.0577 [hep-ph]].

\bibitem{Pasechnik:2013bxa} 
  R.~Pasechnik, V.~Beylin, V.~Kuksa and G.~Vereshkov,
  Phys.\ Rev.\ D {\bf 88}, 075009 (2013)
  [arXiv:1304.2081 [hep-ph]].

\bibitem{Pasechnik:2013kya} 
  R.~Pasechnik, V.~Beylin, V.~Kuksa and G.~Vereshkov,
  Eur.\ Phys.\ J.\ C {\bf 74}, 2728 (2014)
  [arXiv:1308.6625 [hep-ph]].

\bibitem{Lebiedowicz:2013fta} 
  P.~Lebiedowicz, R.~Pasechnik and A.~Szczurek,
  Nucl.\ Phys.\ B {\bf 881}, 288 (2014)
  [arXiv:1309.7300 [hep-ph]].

\bibitem{Hietanen:2014xca} 
  A.~Hietanen, R.~Lewis, C.~Pica and F.~Sannino,
  arXiv:1404.2794 [hep-lat].

\bibitem{Cacciapaglia:2014uja} 
  G.~Cacciapaglia and F.~Sannino,
  arXiv:1402.0233 [hep-ph].

\bibitem{Buckley:2012ky} 
  M.~R.~Buckley and E.~T.~Neil,
  Phys.\ Rev.\ D {\bf 87}, no. 4, 043510 (2013)
  [arXiv:1209.6054 [hep-ph]].

\bibitem{Nussinov}
  S.~Nussinov,
  Phys.\ Lett.\ B {\bf 165}, 55 (1985);\\
  S.~M.~Barr, R.~S.~Chivukula, and E.~Farhi, Phys. Lett. B {\bf 241}, 387 (1990).

\bibitem{Gudnason-1}
  S.~B.~Gudnason, C.~Kouvaris and F.~Sannino,
  Phys.\ Rev.\ D {\bf 74}, 095008 (2006)
  [hep-ph/0608055].

\bibitem{Gudnason-2}
  S.~B.~Gudnason, C.~Kouvaris and F.~Sannino,
  Phys.\ Rev.\ D {\bf 73}, 115003 (2006)
  [hep-ph/0603014].

\bibitem{Khlopov-1}
  M.~Y.~Khlopov and C.~Kouvaris,
  Phys.\ Rev.\ D {\bf 78}, 065040 (2008)
  [arXiv:0806.1191 [astro-ph]].

\bibitem{Khlopov-2}
  M.~Y.~Khlopov, A.~G.~Mayorov and E.~Y.~Soldatov,
  Int.\ J.\ Mod.\ Phys.\ D {\bf 19}, 1385 (2010)
  [arXiv:1003.1144 [astro-ph.CO]].

\bibitem{DelNobile:2011uf}
  E.~Del Nobile and F.~Sannino,
  Int.\ J.\ Mod.\ Phys.\ A {\bf 27}, 1250065 (2012)
  [arXiv:1102.3116 [hep-ph]].

\bibitem{Sannino:2009za}
  F.~Sannino,
  Acta Phys.\ Polon.\ B {\bf 40}, 3533 (2009)
  [arXiv:0911.0931 [hep-ph]].

\bibitem{Ryttov}
  T.~A.~Ryttov and F.~Sannino,
  Phys.\ Rev.\ D {\bf 78}, 115010 (2008)
  [arXiv:0809.0713 [hep-ph]].

\bibitem{Lewis:2011zb}
  R.~Lewis, C.~Pica and F.~Sannino,
  Phys.\ Rev.\ D {\bf 85}, 014504 (2012)
  [arXiv:1109.3513 [hep-ph]].
  
\bibitem{Appelquist:2014dja} 
  T.~Appelquist, E.~Berkowitz, R.~C.~Brower, M.~I.~Buchoff, G.~T.~Fleming, J.~Kiskis, G.~D.~Kribs and M.~Lin {\it et al.},
  arXiv:1402.6656 [hep-lat].

\bibitem{Belyaev:2010kp}
  A.~Belyaev, M.~T.~Frandsen, S.~Sarkar and F.~Sannino,
  Phys.\ Rev.\ D {\bf 83}, 015007 (2011)
  [arXiv:1007.4839 [hep-ph]].

\bibitem{Petraki:2013wwa}
  K.~Petraki and R.~R.~Volkas,
  arXiv:1305.4939 [hep-ph].

\bibitem{Kouvaris:2013gya} 
  C.~Kouvaris,
  Phys.\ Rev.\ D {\bf 88}, no. 1, 015001 (2013)
  [arXiv:1304.7476 [hep-ph]].

\bibitem{Aprile:2012nq}
  E.~Aprile {\it et al.}  [XENON100 Collaboration],
  Phys.\ Rev.\ Lett.\  {\bf 109}, 181301 (2012)
  [arXiv:1207.5988 [astro-ph.CO]].

\bibitem{Akerib:2013tjd} 
  D.~S.~Akerib {\it et al.}  [LUX Collaboration],
  Phys.\ Rev.\ Lett.\  {\bf 112}, 091303 (2014)
  [arXiv:1310.8214 [astro-ph.CO]].

\bibitem{Hisano:2011cs} 
  J.~Hisano, K.~Ishiwata, N.~Nagata and T.~Takesako,
  JHEP {\bf 1107}, 005 (2011)
  [arXiv:1104.0228 [hep-ph]].

\bibitem{Steigman:2012nb} 
  G.~Steigman, B.~Dasgupta and J.~F.~Beacom,
  Phys.\ Rev.\ D {\bf 86}, 023506 (2012)
  [arXiv:1204.3622 [hep-ph]].

\bibitem{Hinshaw:2012aka} 
  G.~Hinshaw {\it et al.}  [WMAP Collaboration],
  Astrophys.\ J.\ Suppl.\  {\bf 208}, 19 (2013)
  [arXiv:1212.5226 [astro-ph.CO]].

\bibitem{Spergel:1999mh} 
  D.~N.~Spergel and P.~J.~Steinhardt,
  Phys.\ Rev.\ Lett.\  {\bf 84}, 3760 (2000)
  [astro-ph/9909386].

\bibitem{Ackermann:2013uma} 
  M.~Ackermann {\it et al.}  [Fermi-LAT Collaboration],
  Phys.\ Rev.\ D {\bf 88}, 082002 (2013)
  [arXiv:1305.5597 [astro-ph.HE]].

\bibitem{Shifman:1978zn} 
  M.~A.~Shifman, A.~I.~Vainshtein and V.~I.~Zakharov,
  Phys.\ Lett.\ B {\bf 78}, 443 (1978).

\bibitem{Giedt:2009mr} 
  J.~Giedt, A.~W.~Thomas and R.~D.~Young,
  Phys.\ Rev.\ Lett.\  {\bf 103}, 201802 (2009)
  [arXiv:0907.4177 [hep-ph]].

\bibitem{Hill:2011be} 
  R.~J.~Hill and M.~P.~Solon,
  Phys.\ Lett.\ B {\bf 707}, 539 (2012)
  [arXiv:1111.0016 [hep-ph]].

\bibitem{Alloul:2013bka} 
  A.~Alloul, N.~D.~Christensen, C.~Degrande, C.~Duhr and B.~Fuks,
  Comput.\ Phys.\ Commun.\  {\bf 185}, 2250 (2014)
  [arXiv:1310.1921 [hep-ph]].

\bibitem{Hahn:2000kx} 
  T.~Hahn,
  Comput.\ Phys.\ Commun.\  {\bf 140}, 418 (2001)
  [hep-ph/0012260].

\bibitem{Hahn:1998yk} 
  T.~Hahn and M.~Perez-Victoria,
  Comput.\ Phys.\ Commun.\  {\bf 118}, 153 (1999)
  [hep-ph/9807565].

\end{thebibliography}
\end{document}